\documentclass[reprint,amsmath,amssymb,aps]{revtex4-1}

\usepackage{graphicx}
\usepackage{dcolumn}
\usepackage{bm}
\usepackage{lmodern}                         
\usepackage[T1]{fontenc}                   
\usepackage{ae}   
\usepackage[utf8]{inputenc}    
\usepackage{xfrac}  
\usepackage{physics}
\usepackage{palatino}
\RequirePackage{dsfont}  
\DeclareMathOperator{\arctanh}{arctanh}           

\begin{document}

%%%%%%%%%%%%%%%%%%%%%%%%%%%%%%%%%%%%%%%%%%%%%%%%%%
\title{Experimental quantum thermodynamics with linear optics}

\author{G. L. Zanin}
\author{T. H\"{a}ffner}
\author{M. A. A. Talarico}
\altaffiliation[Present address: ]{Universidade Tecnol\'{o}gica Federal do Paran\'{a}, CEP 85902-490, Toledo, PR, Brazil}
\author{E. I. Duzzioni}
\author{P. H. Souto Ribeiro}
\email{p.h.s.ribeiro@ufsc.br}
\affiliation{Departamento de F\'{i}sica, Universidade Federal de Santa Catarina, CEP 88040-900, Florian\'{o}plis, SC, Brazil}
\author{G. T. Landi}
\affiliation{Instituto de F\'{i}sica, Universidade de S\~{a}o Paulo, 05314-970 S\~{a}o Paulo, S\~{a}o Paulo, Brazil}
\author{L. C. C\'{e}leri}
\email{lucas@chibebe.org}
\affiliation{Institute of Physics, Federal University of Goi\'{a}s, 74690-900, Goi\^{a}nia, GO, Brazil}

\begin{abstract}
The study of non-equilibrium physics from the perspective of the quantum limits of thermodynamics and fluctuation relations can be experimentally addressed with linear optical systems. We discuss recent experimental investigations in this scenario and present new proposed schemes and the potential advances they could bring to the field.
\end{abstract}

\maketitle

%%%%%%%%%%%%%%%%%%%%%%%%%%%%%%%%%%%%%%%%%%%%%%%%%%%%%%%%%%%
\section{Introduction}
\label{sec:intro}

Experiments lie at the heart of all natural sciences. Despite the great success achieved by thermodynamics since the industrial revolution, its experimental investigation can still bring important advances, not only from the fundamental point of view (to test the limits of its applicability) but also for practical purposes towards new technologies. 

Since Carnot, Clausius, Maxwell, Boltzmann, Gibbs and others constructed its basis more than one hundred years ago, thermodynamics witnessed a huge development, passing through many conceptual shifts. It was initially developed as a macroscopic theory, aiming to describe very specific measurements consisting of spatial and temporal averages. The advent of statistical mechanics and quantum theory pushed thermodynamics to a higher level. Among several developments, we can mention Onsager's theory \cite{onsagerI,onsagerII}, Kubo's fluctuation-dissipation theorem \cite{kubo} and the generalized fluctuation relations derived by Jarzynski \cite{jar1997}, Crooks \cite{crooks} and others \cite{Camhantal01,Camhantal02}.

These tools were developed in order to understand how the laws of thermodynamics apply to small (classical and quantum) systems, where fluctuations matter. As a consequence, it was also necessary to develop experimental techniques able to probe such limits. Regarding quantum systems, the requirement of performing two energy projective measurements on the system, for testing Jarzynski fluctuation relation, creates a huge barrier for experimental investigations. This fact explains why we have a relatively small number of reported experiments to date. For instance, pioneering experiments were performed for studying Jarzynski and crooks relations \cite{szabo,bustamante2002,bustamante2005,bechinger,pekola,kiang,sano}. There were also experimental investigations of the Landauer's principle \cite{snider,lutz,bechhoefer} and Maxwell's demon paradox \cite{averin,Barbieri16}. The experiments for quantum systems employed several platforms. Jarzynski's equality and Landauer's principle for quantum systems were addressed using nuclear spins \cite{batalhao2014,celeri}. They used a strategy based on Ramsey interferometry for avoiding the energy projective measurements \cite{vedral,paternostro}. The so called two-point measurement protocol to test Jarzynski's relation was implemented using ion-trap \cite{jarion} and an all-optical set-up \cite{Araujo18}. A trapped-ion setup was also employed to implement a quantum thermal machine \cite{Rossnagel325}. Finally, Maxwell's demon paradox was addressed in a superconducting-device experiment \cite{huard}. 

Our aim here is to present a brief review of all-optical platforms and describe how this kind of setup can be employed for studying quantum thermodynamics. Motivated by the reliability of this experimental approach, we also present new theoretical results that can be implemented with them.

The paper is organized as follows. Section \ref{sec:work} is devoted to the discussion of some aspects of the second law of thermodynamics and fluctuation relations. It is not our intention here to provide a complete review of such topic, but instead we concentrate ourselves in the main aspects that will be important for the experimental investigations to be presented later. In short, we discuss the statistical character of the second law when we consider small (quantum) systems and how Jarzynski equality emerges from it. In order to do this, we consider the two-point measurement definition of work (see Ref. \cite{rahav} and references therein). The notions of entropy production and irreversibility are also discussed in this section. In Sec. \ref{sec:analogy}, the isomorphism between the paraxial wave equation and the two dimensional Sch\"{o}dinger equation is presented. This mapping allows us to study the thermodynamics of quantum systems employing all-optical experiments. In Sec. \ref{sec:theo} we discuss two theoretical proposals that are suitable for implementation with optical setups. Section \ref{sec:exp} is devoted to the discussion of three experiments that illustrate the utility of this setup. The first one concerns thermometry, the second one being a proof-of-principle for realizing a photonic Maxwell's demon, while the third describes the reconstruction of the work probability distribution for a quantum system. Conclusions and perspectives are presented in Sec. \ref{sec:conc}.

Throughout the article we use units such that Boltzmann and Plank constants are equal to one. 

%%%%%%%%%%%%%%%%%%%%%%%%%%%%%%%%%%%%%%%%%%%%%%%%%%%%%%%%%%%%
\section{The second law in quantum and classical thermodynamics}
\label{sec:work}

Different from fundamental laws of physics, like Newton's or Maxwell's equations, the second law of thermodynamics sets limits for all physical process. Its importance is not only practical (setting the efficiency of heat engines, for instance), but also fundamental, since it tells us the preferred direction of time (the so called arrow of time). Despite its universal character, in the sense that its formulation is independent of any microscopic details of the considered system, there are deep conceptual differences between quantum and classical descriptions. The goal of this section is to shortly review these ideas.

%%%%%%%%%%%%%%%%%%%%%%%%%%%%%%%%%%%%%%%%%%%%%%%%%%%%%%%%
\subsection{Jarzynski equality}
\label{ssec:work_Jar}

The idea of work extraction is among the most important in thermodynamics \cite{kondeprigo, Jarzynski-rev, callen}. It is the basic figure of merit dictating the construction of heat engines and related devices. The limitations imposed on it by the second law of thermodynamics reflect some of the deepest ideas in physics. According to the second law, the amount of work $\mathcal{W}$ that must be invested in order to perform a physical process is lower bounded by 
\begin{equation}\label{work_2nd}
\mathcal{W} \geq \Delta F,    
\end{equation}
where $\Delta F = F_{\tau} - F_{0}$ is the change in free energy $F = U - TS$, with $U$ being the internal energy, $T$ the temperature and $S$ the entropy. The considered process is assumed to take place in the time interval $t \in [0,\tau]$. In Eq.~(\ref{work_2nd}) work is defined to be negative when it is extracted (that is, when the system performs work on an external agent). Thus, for work extraction Eq.~(\ref{work_2nd}) should be read as $\vert\mathcal{W}\vert \leq \vert\Delta F\vert$. We therefore see that $F$ is the energy that is \emph{free} to be potentially extracted as useful work. However, in general not all invested energy translates into free energy (that can be converted into useful work), as some energy may be irreversibly dissipated. This fundamental limitation on the amount of work that can be extracted, or the minimal amount of work that must be invested to increase free energy, is the essence of the second law of thermodynamics.

For more than a century, thermodynamics has been restricted to macroscopic systems. In the last two decades, however, novel formulations appropriate for the microscopic realm have been introduced, which led to an increasing interest in the physics community. All these formulations rely on a fundamental paradigm shift, namely, that in order to properly address
the thermodynamics of microscopic systems, one must take into account \emph{fluctuations} in physical quantities, like work for instance. In the micro-world fluctuations play a prominent role, so that work and all other thermodynamic quantities will also fluctuate, being therefore described by random variables. One may then speak of a distribution of work, $P(\mathcal{W})$, which gives the probability that a certain amount of work $\mathcal{W}$ is extracted in a single run of a process. We will focus on work, but the ideas presented here can be readily extended to other thermodynamic quantities.

Stochastic thermodynamics is the research field that treats the classical contribution from such fluctuations, whose origin lies on thermal effects \cite{sekimoto}. However, in sufficiently well controlled systems, fluctuations may also have a quantum contribution, thus opening the possibility of exploiting genuinely quantum mechanical resources to perform thermodynamic tasks with unprecedented efficiency \cite{Rousselet1994, PhysRevLett741504, sano, jarion, batalhao2014, Rossnagel325}. This fascinating new perspective is the main motivation behind the blooming field of quantum thermodynamics.

In addition to the potential technological implications, stochastic and quantum thermodynamics also provides valuable insight into the second law. More specifically, on how the intrinsically irreversible behavior of macroscopic systems ultimately emerges from the underlying reversible dynamics of the microscopic constituents. Perhaps the most dramatic manifestations of this aspect are the so called fluctuation relations, whose most famous representative is Jarzynski's equality. It was first derived for classical systems \cite{jar1997} and then extended to the quantum realm \cite{Tasaki, kurchan2001, mukamel}. It reads 
\begin{equation}\label{eq:01Jar}
\left\langle e^{-\beta\mathcal{W}}\right\rangle = e^{-\beta\Delta F},
\end{equation}
where $\beta=1/T$ is the inverse temperature and
\begin{equation}\label{work_average_PW}
\left\langle e^{-\beta\mathcal{W}}\right\rangle = \int d \mathcal{W}\; P(\mathcal{W}) e^{-\beta \mathcal{W}}.
\end{equation}
There are several remarkable aspects of Eq.~(\ref{eq:01Jar}). First, it is an equality, even though it is valid for processes \emph{arbitrarily} far from equilibrium. This is in stark contrast to equilibrium thermodynamics, which is only capable of offering inequalities for non-equilibrium processes. Second, equilibrium information (the free energy) is fundamentally encoded into the response of the system. Finally, the derivation of Eq.~(\ref{eq:01Jar}) relies only on the assumption that the initial state is thermal and the underlying dynamics (e.g. Newton's law or Schr\"{o}dinger's equation) is time-reversal invariant. This hints at the universality of non-equilibrium processes.

Using Jensen's inequality in Eq.~(\ref{eq:01Jar}) one concludes that 
\begin{equation}\label{work_jensen}
\langle \mathcal{W}\rangle \ge \Delta F.
\end{equation} 
We therefore recover the traditional second law (\ref{work_2nd}), but  for the average work $\langle \mathcal{W} \rangle$ instead. Individual realizations of a work process may violate Eq.~(\ref{work_2nd}), but Eq.~(\ref{work_jensen}) should always hold. This fact points out the statistical character of the second law of thermodynamics. As a consequence of the central limit theorem, fluctuations must vanish in the thermodynamic limit (large systems), thus implying that the work distribution $P(\mathcal{W})$ should become more and more peaked around the average value $\langle \mathcal{W} \rangle$. Therefore, for macroscopic systems, local violations of Eq.~(\ref{work_2nd}) become exponentially less likely. In this way, classical thermodynamics is recovered in the macroscopic limit. 

%Work%%%%%%%%%%%%%%%%%%%%%%%%%%%%%%%%%%%%%%%%%%%%%%%%%%%%%%%%%%%
\subsection{Work distribution in quantum mechanical systems}
\label{ssec:work_dist}

In this paper we shall be concerned with the work distribution for quantum systems undergoing a unitary work process. In this case the distribution of work may be constructed using the two-point measurement protocol \cite{mukamel}, which goes as follows. 
\begin{itemize}
    \item The system, whose Hamiltonian is $H_0$, is prepared in thermal equilibrium at temperature $T$
    \[
    \rho_{0}^\text{th} = \frac{\text{e}^{-\beta H_0}}{Z_0},
    \]
    where $Z_0 = \text{tr}(e^{-\beta H_0})$ is the partition function.
    \item After this, a projective energy measurement is performed on the system. State $|\varepsilon_n^0\rangle$ will be found with probability 
\begin{equation}
p_n^0=\frac{e^{-\beta \varepsilon^{0}_{n}}}{Z_0}.
\label{eq:thermal}
\end{equation}
We defined the eigenvalues and eigenvectors of the initial Hamiltonian as $H_0 |\varepsilon_n^0\rangle = \varepsilon_n^0 |\varepsilon_n^0\rangle$. 
    \item The next step is the process (work protocol), which is characterized by an externally controlled parameter $\lambda_t$ (or set of parameters). This unitary process changes the Hamiltonian from $H_0$ to a final value $H_{\tau}$. The process is denoted by $\mathbf{U}_{\tau}$.
    \item The final step is a projective energy measurement on the final Hamiltonian eigenbasis, defined by  $H_{\tau} |\varepsilon_m^{\tau}\rangle = \varepsilon_m^{\tau} |\varepsilon_m^{\tau}\rangle$. State $|\varepsilon_m^{\tau}\rangle$ is found with probability 
\begin{equation}
p_{m|n} = |\langle \varepsilon^{\tau}_{m}| \mathbf{U}_{\tau} |\varepsilon^{0}_{n}\rangle|^{2}.
\label{eq:prob-work}
\end{equation}
The sequence of quantum numbers $(n,m)$ forms the quantum trajectory for this process, which occurs with  path probability 
\begin{equation}
    p_{m,n} = p_{n}^{0}p_{m|n}.
    \label{eq:totalprob}
\end{equation}
\end{itemize}

It is important to observe that the system is assumed to be decoupled from any environments during the time window $t\in[0,\tau]$ where the work protocol is implemented. Consequently, the work performed in each trajectory will simply be defined as the change in the energy of the system (sometimes referred to as inclusive work \cite{jar2007}) 
\begin{equation}
\mathcal{W}_{m,n} = \varepsilon^{\tau}_{m} - \varepsilon^{0}_{n}.
\label{eq:jar01}
\end{equation}
The probability distribution of work may then be computed from the general definition
\begin{equation}
P(\mathcal{W}) = \sum\limits_{m,n} \delta \left( \mathcal{W} - \mathcal{W}_{m,n}\right)p_{m,n}.
\label{eq:rel11}
\end{equation}
From $P(\mathcal{W})$ all statistical quantities can be computed in the usual way. For instance, the average work is nothing but
\begin{equation}
    \langle \mathcal{W} \rangle
    = \int d\mathcal{W} \; P(\mathcal{W}) \mathcal{W} 
    =\sum_{m,n} p_{m,n}\mathcal{W}_{m,n}.
\end{equation}
With some rearrangements, one may show that this can also be written as 
\begin{equation}
    \langle \mathcal{W} \rangle = \text{tr}\big\{ H_{\tau} \rho_{\tau}\big\} -\text{tr}\big\{H_0 \rho_0\big\},
\end{equation}
where $\rho_{\tau} = U_{\tau} \rho_0 U_{\tau}^\dagger$ is the final state of the system when measurements are suppressed. Unfortunately, extending this reasoning to higher order moments of $\mathcal{W}$ is not possible, as work is not a function of state and therefore cannot be associated with a quantum mechanical observable \cite{talkner}. Consequently, higher order moments are only correctly defined using the two-point measurement protocol and the corresponding distribution $P(\mathcal{W})$. 

%%%%%%%%%%%%%%%%%%%%%%%%%%%%%%%%%%%%%%%%%%%%%%%%%%%%%%%%%%%%
\subsection{Irreversibility and entropy production}
\label{ssec:irrevers}

The second law in Eq.~(\ref{work_jensen}) (or its macroscopic counterpart in Eq.~(\ref{work_2nd})) reflects the intrinsically irreversible nature of a physical process. We define the irreversible work as 
\begin{equation}\label{irr_Wirr}
    \mathcal{W}_\text{irr} = \langle \mathcal{W} \rangle - \Delta F \geq 0,
\end{equation}
which quantifies the amount of free energy that was not harnessed as useful work. The process will be said reversible when $\mathcal{W}_\text{irr} = 0$, which occurs for quasi-static transformations. In this case the energy invested in order to perform the process is entirely converted into free energy.

The irreversible work can be viewed as a specific manifestation of a more general concept in thermodynamics known as \emph{entropy production}. The average entropy production associated with the irreversible work (\ref{irr_Wirr}) is defined as 
\begin{equation}\label{irr_sigma}
\Sigma = \beta \mathcal{W}_\text{irr} = \beta (\langle \mathcal{W} \rangle - \Delta F).
\end{equation}
The concept of entropy production was introduced by Clausius as a quantifier of irreversibility for general thermodynamic processes. The specific definition of $\Sigma$ depends on the process in question. Notwithstanding, the basic idea is that in terms of the entropy production one may formulate the second law as a single universal expression
\begin{equation}\label{irr_2nd}
\Sigma \geq 0.
\end{equation}
Despite its simplicity and elegance, this is perhaps the most important expression in all of thermodynamics and certainly the one with the deepest conceptual implications.

Following the same reasoning outlined in the last section, for small (quantum) systems one may also define a fluctuating entropy production as 
\begin{equation}
    \sigma_{m,n} = \beta(\mathcal{W}_{m,n} - \Delta F),
\end{equation}
where $\mathcal{W}_{m,n}$ is given in Eq. (\ref{eq:jar01}). The average entropy production (\ref{irr_sigma}) is then recovered as $\Sigma = \langle \sigma \rangle$. Moreover, Jarzynski equality (\ref{eq:01Jar}) may now be rewritten in the form of a \emph{fluctuation theorem}
\begin{equation}\label{irr_FT}
    \langle e^{-\sigma}\rangle = 1.
\end{equation}
This expression can be viewed as a universal result for entropy production, since it is independent of the considered physical process. As we saw before, it also encompasses the second law (\ref{irr_2nd}), since the application of Jensen's inequality to Eq. (\ref{irr_FT}) immediately leads to Eq. (\ref{irr_2nd}).

\begin{figure}[!h]
\centering
\includegraphics[width=0.42\textwidth]{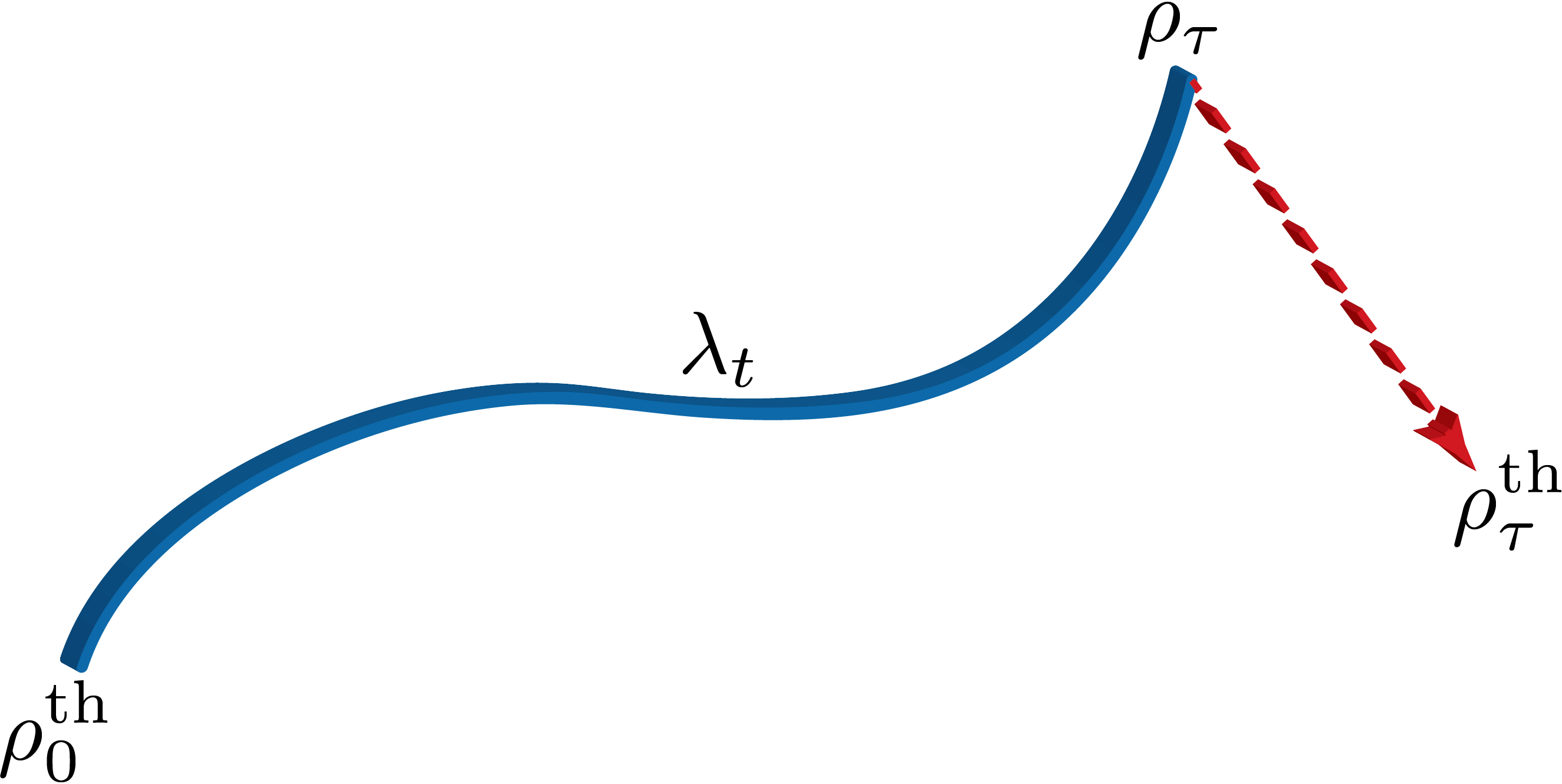}
\caption{Diagram of the basic dynamics undergone by the system during a work protocol. The system is initially prepared in a thermal state $\rho_0^\text{th}$ with Hamiltonian $H_0$. After the driving protocol, the system will be in a non-equilibrium state $\rho_{\tau}$, which will in general be different from the thermal one, defined by $\rho_{\tau}^\text{th} = e^{-\beta H_{\tau}}/{Z_{\tau}}$.}
\label{fig:diagram_entropy_production}
\end{figure}

Next, let us return to Eq. (\ref{irr_sigma}) and the unitary driving scenario discussed earlier. A diagram of the dynamics is shown in Fig. \ref{fig:diagram_entropy_production}. Even though the system was initially in a thermal equilibrium state, due to the action of the driving protocol, the final state $\rho_{\tau}$ will in general be a non-equilibrium one. After some calculations, the entropy production may also be written as \cite{kuwai}
\begin{equation}\label{irr_sigma_KL}
\Sigma = S(\rho_F || \rho_{\tau}^\text{th}),
\end{equation}
where $S(\rho||\sigma) = \tr(\rho \ln \rho - \rho \ln \sigma)$ is the quantum relative entropy and $\rho_{\tau}^\text{th}$ is the thermal state associated with the final Hamiltonian. Hence, we see that the entropy production may also be interpreted in terms of how distinguishable the actual final state $\rho_{\tau}$ is from the reference thermal state $\rho_{\tau}^\text{th}$.

%%%%%%%%%%%%%%%%%%%%%%%%%%%%%%%%%%%%%%%%%%%%%%%%%%%%%%%%%%%%
%%%%%%%%%%%%%%%%%%%%%%%%%%%%%%%%%%%%%%%%%%%%%%%%%%%%%%%%%%%

Now we have the basic ingredients of thermodynamics of small systems. We next introduce a platform for the experimental investigation of such ideas on an optical setup.

%%%%%%%%%%%%%%%%%%%%%%%%%%%%%%%%%%%%%%%%%%%%%%%%%%%%%%%%%%%%%%
\section{Analogy between the paraxial wave equation and the 2D Schr\"{o}dinger equation}
\label{sec:analogy}

The paraxial wave equation describes light beams that do not diverge (or converge) too much during propagation. In this section we review the isomorphism between this equation and the two-dimensional Schr\"{o}dinger equation. We start from the Helmholtz equation for a light field
\begin{equation}
(\nabla^2 + k^2) A(x,y,z)  = 0,
\label{eq:parax1}
\end{equation}
where $A(x,y,z)$ describes the spatial dependence for the amplitude of the electric field. Denoting by $k_0 = 2\pi/\lambda_0$ the wavenumber for a medium with constant index of refraction $n_0$, let us suppose that $A(x,y,z)$ can be written as $A = u(x,y,z)\mbox{e}^{-ik_0z}$, where $z$ is the 
direction of propagation. This starting point includes the assumption that the field is almost monochromatic, so that the phase $\exp{-i \omega_{opt} t}$ is factored out. Inserting the amplitude $A(x,y,z)$ in the wave equation, and employing the paraxial approximation 
\begin{equation}
\left| \frac{\partial^2 u}{\partial z^2}\right| \ll \left|k_0 \frac{\partial u}{\partial z}\right|,
\label{eq:parax2}
\end{equation}
we get  
\begin{equation}
\nabla^2_{\perp} u(x,y,z) - i k_0 \frac{\partial u(x,y,z)}{\partial z} = 0,
\label{eq:parax3}
\end{equation}
where $\nabla^2_{\perp} = \partial^2/\partial x^{2} + \partial^2/\partial y^{2}$. This equation is known as the paraxial Helmholtz equation, or simply paraxial wave equation. It can be written in a more convenient form
\begin{equation}
\frac{i}{k_0} \frac{\partial u(x,y,z)}{\partial z} = - \frac{1}{k_0^2}\nabla^2_{\perp} u(x,y,z) ,
\label{eq:parax3}
\end{equation}
where the minus sign in the right hand side comes from the arbitrary definition of the sense of propagation $\pm z$. 
We can directly connect it to the Sch\"{o}dinger equation by making the identifications
\begin{eqnarray}
\nonumber
&\psi(x,y,t)& \rightarrow u(x,y,z), \\ \nonumber
&t& \rightarrow z , \\ \nonumber
&\hbar& \rightarrow \frac{1}{k_0} ,
\label{eq:correspondence}
\end{eqnarray}
and comparing it to the Scr\"{o}dinger equation for the free particle
\begin{equation}
i \hbar \frac{\partial \psi(x,y,t)}{\partial t} =  -\frac{\hbar^2}{2m} 
\nabla^2_{\perp}  \psi(x,y,t).
\label{eq:schrodinger}
\end{equation}

This equivalence can be extended to the case where the Scr\"{o}dinger equation describes a particle under the action of some potential. In this case, the wave equation should be solved in a non-homogeneous medium with a position dependent index of refraction $n(x,y)$. Rigorously speaking, one should come back to the Maxwell's equations and not directly employ the wave equation. However, under the paraxial approximation, and considering that the variation of the index of refraction with $x$ and $y$ is small enough, it is possible to use the same reasoning employed above including a position dependent potential. A complete discussion about this subject can be found in Ref. \cite{Marcuse}, and the so called {\em optical Schr\"{o}dinger equation} is derived in Ref. \cite{Marte}, for instance. Based on these works, we find that, when a potential for the particle is included, we can write
\begin{equation}
\frac{i}{k_0} \frac{\partial u(x,y,t)}{\partial z} = - \frac{1}{k_0^2}
\left(\nabla^2_{\perp} +  n(x,y) \right) u(x,y,t).
\label{eq:paraxialpotential}
\end{equation}
where $n(x,y)$ is a position-dependent index of refraction, and the corresponding Scr\"{o}dinger equation is
\begin{equation}
i \hbar \frac{\partial \psi(x,y,t)}{\partial t} =  -\frac{\hbar^2}{2m} 
\left( \nabla^2_{\perp}  + V(x,y) \right)  \psi(x,y,t).
\label{eq:schrodinger}
\end{equation}

In this context, it is convenient to make $m=1/2$ (without loss of generality) since the mass has no physical meaning in the optical case. 

%%%%%%%%%%%%%%%%%%%%%%%%%%%%%%%%%%%%%%%%%%%%%%%%%%%%%%%%%%%%%
\begin{figure}[!h]
	\includegraphics[width=0.42\textwidth]{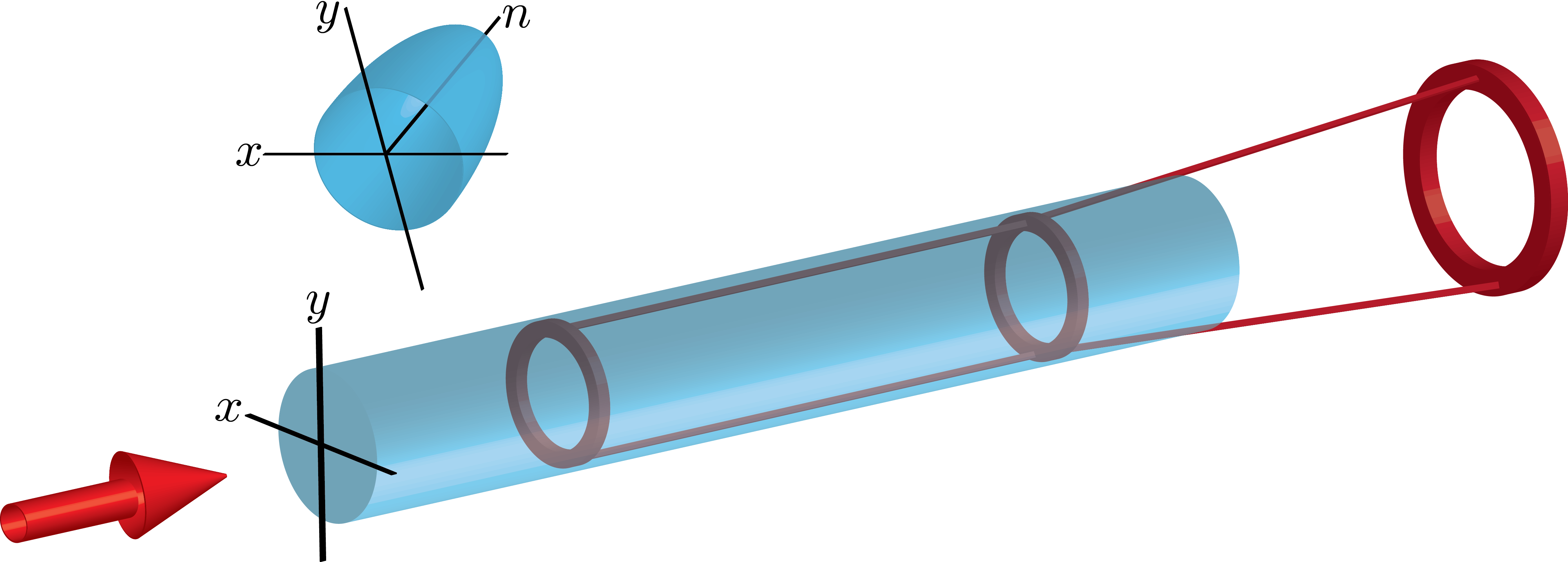}
	\caption{A light source shines a transparent rod with transverse modulation of the index of refraction given by a revolution paraboloid. Only compatible modes like the Laguerre-Gaussian ones are coupled.}
	\label{fig:paraxial}
\end{figure}
%%%%%%%%%%%%%%%%%%%%%%%%%%%%%%%%%%%%%%%%%%%%%%%%%%%%%%%%%%%%%

An important potential function is the quantum harmonic oscillator (QHO), which we explore in the context of quantum thermodynamics. The index of refraction that corresponds to the QHO potential $V(x,y) = \omega^2 (x^2 +y^2)/2$ is given by
\begin{equation}
 n(x,y) = \frac{1}{2} n\alpha (x^2 +y^2) - n^2,
\label{eq:qho}
\end{equation}
where $\omega$ is the angular frequency of the QHO, $n$ is the index of refraction in the center ($x$ = 0, $y$ = 0) of the propagation medium and $\alpha$ is a constant. This index function describes the so called {\em square law media} that appears in the context of optical waveguides \cite{Marcuse}. The energy of the QHO is related to the optical parameters
by
\begin{equation}
\hbar \omega = \frac{\sqrt{n\alpha}}{k_0}.
\label{eq:energy}
\end{equation}

We can check the consistency of the analogy by noting that the classical limit given by $\hbar \rightarrow 0$ corresponds to $\lambda \rightarrow 0$, which is equivalent to the ray-optics limit. We also verify that increasing $\omega$ corresponds to increasing $\alpha$, which is a parameter that increases the confinement of the light beam through the variation of the refraction index.

\begin{figure}[!h]
	\includegraphics[width=0.42\textwidth]{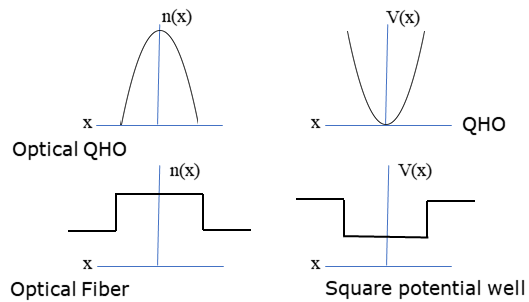}
	\caption{Comparison between the distribution of index of refraction along the direction $x$ with the equivalent potential energy for the cases of an optical fiber/square potential as well as the QHO/optical version.}
	\label{fig:compar}
\end{figure}

In Fig. \ref{fig:paraxial}, the optical analogue of the quantum harmonic oscillator is sketched. The blue rod represents a medium with an index of refraction that varies along the $x$ and $y$ directions according to a paraboloid of revolution $n(x,y)$ given in Eq. \ref{eq:qho}. The index is maximal in the center and decreases when the radius $r^2 = x^2 + y^2$ increases following a parabolic function. In the analogy, this is the optical equivalent of a two-dimension harmonic oscillator. When light shines on the rod, only modes that are solutions to the paraxial wave equation including the index of refraction paraboloid function can propagate inside it. The Laguerre-Gaussian modes, for instance, are such solutions. They are equivalent to the quantum eigenstates, and they will propagate without any change besides a global phase shift. However, outside the rod, the index of refraction is constant. As a result, the beam will diffract and diverge. It is interesting to note that an optical fiber implements the equivalent of a finite height square potential: $n(x,y) = n_c > 1$ for $r = \sqrt{x^2 + y^2} < r_0$ and $n(x,y) =  1$ for $r = \sqrt{x^2 + y^2} > r_0$.  Fig. \ref{fig:compar} shows a comparison between the one dimensional versions of potential $V(x)$ and the modulation $n(x)$ for the square function, which is equivalent to the optical fiber, and the parabolic function for the harmonic oscillator. 

%%%%%%%%%%%%%%%%%%%%%%%%%%%%%%%%%%%%%%%%%%%%%%%%%%%%%%%%%%%% 
%%%%%%%%%%%%%%%%%%%%%%%%%%%%%%%%%%%%%%%%%%%%%%%%%%%%%%%%%%%%

\begin{figure}[!htb]
%\vspace*{0.7cm}
\hspace*{0.5cm}
   \begin{minipage}{0.45\textwidth}
   \vspace*{1cm}
     \frame{\includegraphics[width=.9\linewidth]{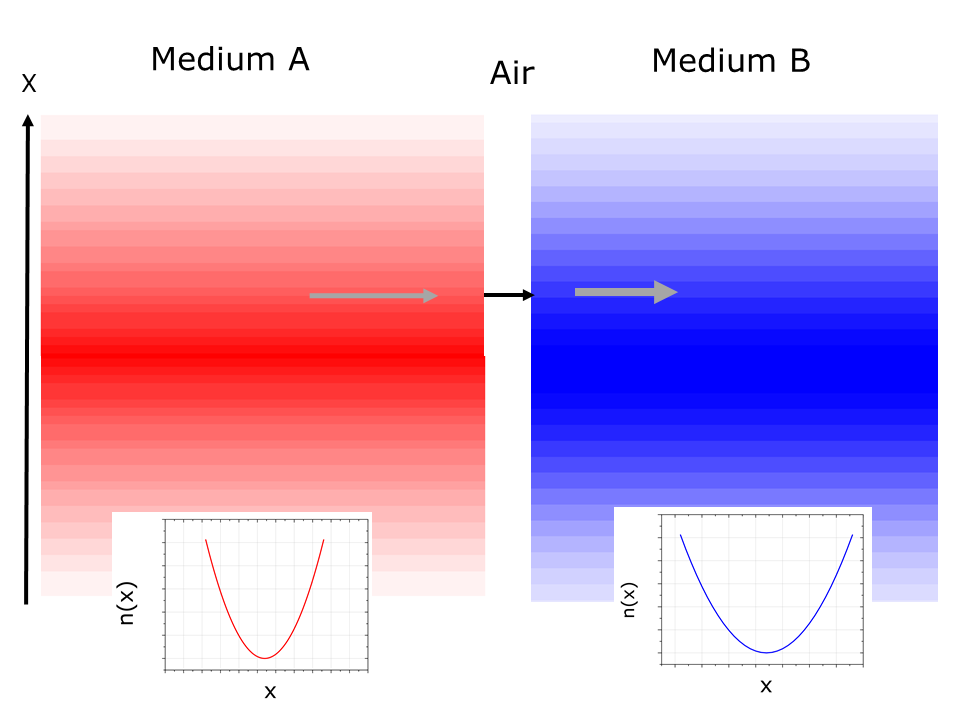}}
%     \caption{1D representation of two square law media with different values of $\alpha$ for a vertically variable. The medium to the left has a narrower  }\label{Figura4a}
   \end{minipage}
   \hspace*{0.5cm}
   \begin {minipage}{0.45\textwidth}
     \frame{\includegraphics[width=.9\linewidth]{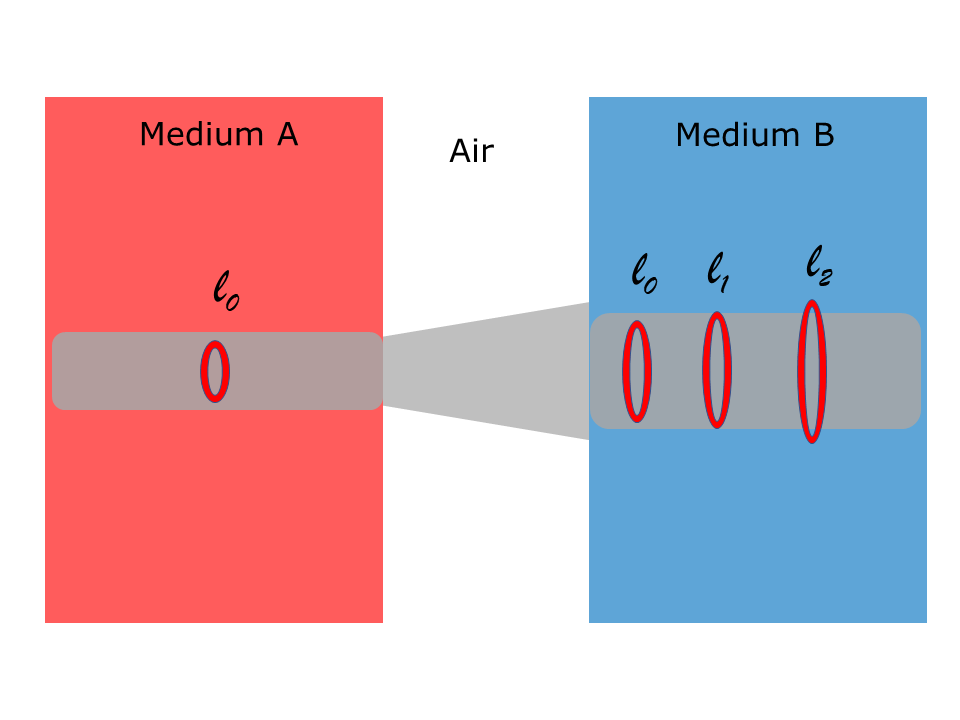}}
     \caption{Upper pannel: 1D representation of two square law media with different values of $\alpha$ defined in Eq. \ref{eq:qho}. Medium A is squeezed
     with respect to Medium B. Suppose there is light propagating from left to right. Lower pannel: graphical representation of an eigenmode of Medium A, a
     Laguerre-Gaussian mode with $\ell = \ell_0$, propagating from left to right. Inside A it only acquires a phase. In the air it diffracts and in B it couples to
      the local eigenmodes $\ell_0, \ell_1, \ell_2$. }
        \label{fig4}
   \end{minipage}
 \end{figure}

Let us now analyze what happens with the optical analogue, when the Hamiltonian of the QHO changes due to some action (process) on the system. The most relevant change concerns some modification in the potential, so that the energy gap $\hbar \omega$ is modified. We refer to this change as a squeezing or anti-squeezing. This optically is accomplished by changing $\alpha$, which means that the paraboloidal index of refraction distribution becomes broader or narrower in the plane $x,y$. The upper pannel of Fig. \ref{fig4} illustrates an anti-squeezig operation represented in one dimension.  Changing $k_0$, or the wavelength, produces a similar effect. The lower pannel of Fig. \ref{fig4} illustrates the effect of changing the propagation medium from A to air and then to Medium B in the light beam. Supposing an initial eigenmode of A, it propagates acquiring a global phase (not represented) and not changing its shape. Reaching the free space, it will diffract and when it is incident opon Medium B, it will couple its energy to eigenmodes od B. This means that the initial beam with OAM $\ell_0$ can give rise to other beams with different values of OAM.

%%%%%%%%%%%%%%%%%%%%%%%%%%%%%%%%%%%%%%%%%%%%%%%%%%%%%%%%%%%%%%%%
%\begin{figure}[!h]
%\includegraphics[width=0.5\textwidth]{Figura4a}
%\includegraphics[width=0.42\textwidth]{Figura4b}
%\caption{A single mode belonging to a family of modes in medium $A$ excites modes belonging to another family of modes of medium $B$, when propagating from one medium to the other. }
%\label{fig:change}
%\end{figure}

%%%%%%%%%%%%%%%%%%%%%%%%%%%%%%%%%%%%%%%%%%%%%%%%%%%%%%%%%%%%%%%
\par
\begin{figure}[!h]
	\includegraphics[width=0.42\textwidth]{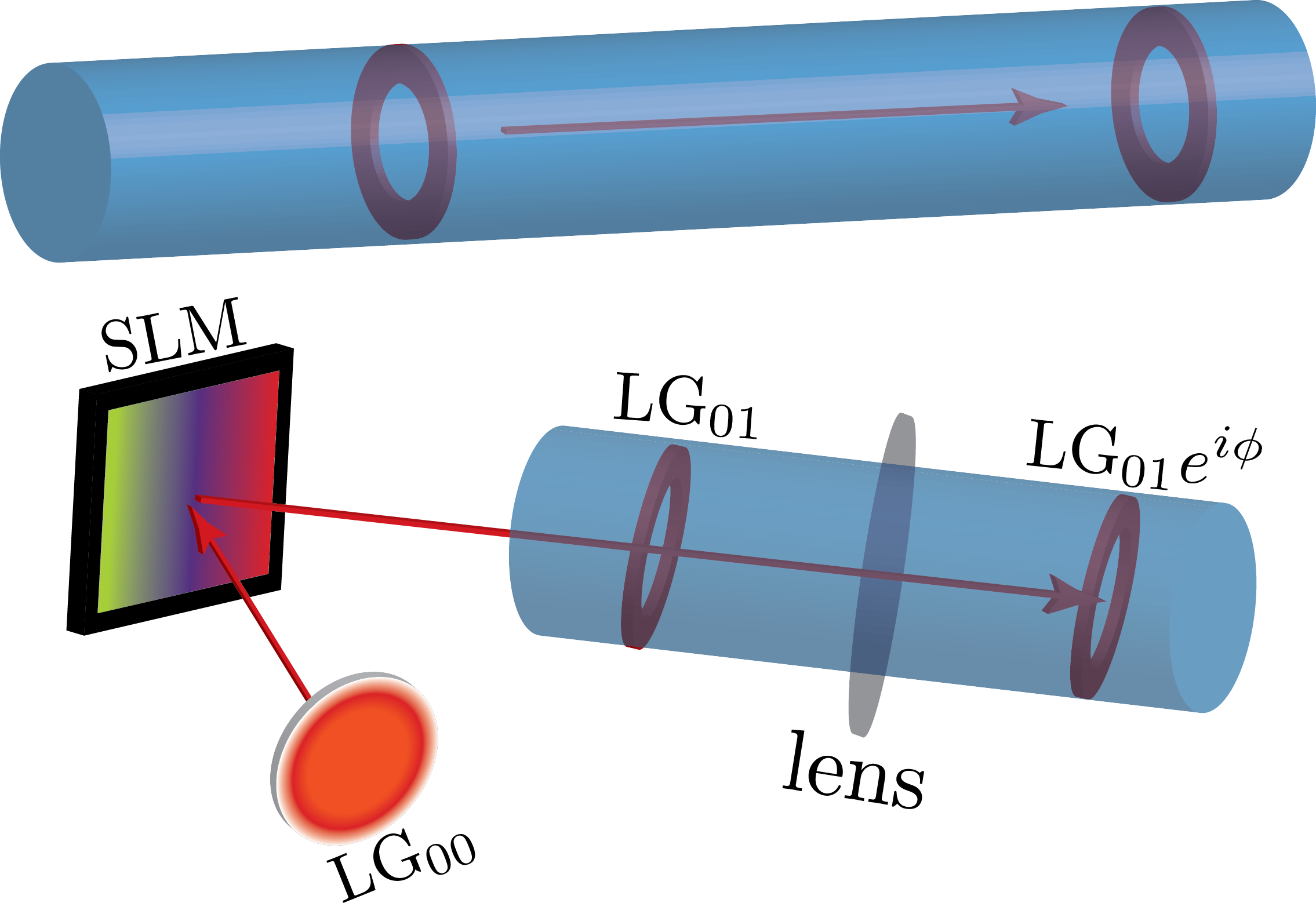}
	\caption{The light mode (eigenmode) propagating in the blue rod can be emulated with a zero order Gaussian beam (LG$_{00}$) incident on a SLM, which prepares a mode identical to the mode inside the rod, and its free propagation inside the rod is realized with free propagation and a lens.}
	\label{fig:stroboscopic}
\end{figure}

In general, it is possible to emulate one quantum particle in an arbitrary potential by using light and a medium where the index of refraction is suitably modulated. However, for practical purposes, it is interesting to replace the modulated medium with a stroboscopic dynamics, where the light beams are actually propagated in free space and then spatially modulated. The time evolution is obtained for a sequence of intermediate plans. Fig. \ref{fig:stroboscopic} illustrates this approach. The light mode inside the modulated medium evolves acquiring only a global phase. The stroboscopic version of this evolution takes the input state and transforms it into the evolved state by means of linear optical components like lenses and spatial light modulators (SLM). 

%%%%%%%%%%%%%%%%%%%%%%%%%%%%%%%%%%%%%%%%%%%%%%%%%%%%%%%%%%%%%
%%%%%%%%%%%%%%%%%%%%%%%%%%%%%%%%%%%%%%%%%%%%%%%%%%%%%%%%%%%

%%%%%%%%%%%%%%%%%%%%%%%%%%%%%%%%%%%%%%%%%%%%%%%%%%%%%%%%%%%%
\section{Theoretical results}
\label{sec:theo}

We start this section presenting the characteristic function approach to investigate the work probability distribution. Next, we discuss experimental schemes considering the quantum harmonic oscillator. Although the results pointed out here are theoretical in nature, they clearly highlight the power of the optical setup in the experimental investigation of quantum thermodynamics.

%%%%%%%%%%%%%%%%%%%%%%%%%%%%%%%%%%%%%%%%%%%%%%%%%%%%%%%%%%%%

\subsection{Work distribution with paraxial light modes: Characteristic function approach}
\label{ssec:thermparax}

The experimental investigation of non-equilibrium behavior of quantum and classical systems becomes increasingly difficult as the size of the system decreases. Measuring the energy states of a quantum system and computing the work distributions becomes problematic, since the two-point measurement protocol for defining the work performed on the system (see Sec. \ref{ssec:work_dist}) requires two energy projective measurements on the system, before and after the process took place. A way of avoiding this difficulty is to reconstruct the work distribution through the characteristic function. The work characteristic function is the Fourier transform of the work distribution and is defined as \cite{batalhao2014}
\begin{eqnarray}
\nonumber 
G(s) =& \displaystyle\int P(\mathcal{W}) e^{is\mathcal{W}} d\mathcal{W} \\
=& ~~\sum\limits_{m,n} p_{m,n}~ \text{e}^{is(\varepsilon^{\tau}_{m} - \varepsilon^{0}_{n}) },
\label{eq:7.1}
\end{eqnarray}
where $P(\mathcal{W})$ is the work probability distribution introduced in Eq. \eqref{eq:rel11}. By measuring the characteristic function, we can reconstruct the work distribution by calculating the inverse Fourier transform. While both the work distribution and the characteristic function are functions of the energy differences between energy levels, or eigenvalues of the system's Hamiltonian, it is possible to encode this information in the phase of an auxiliary system \cite{vedral,paternostro}.

Considering the optical setup, this phase information can be measured at the output of a suitable interferometer in the form of oscillations \cite{Talarico16}. Due to the analogy between the paraxial equation and the Schr\"{o}dinger's equation (see Sec. \ref{sec:analogy}) we can emulate the dynamics of a quantum harmonic oscillator (QHO) with light modes that are solutions to the paraxial equation. These light modes are isomorphic to the energy eigenstates of the QHO \cite{Nienhuis93}. Here, we deal with Hermite-Gaussian (HG) modes, which are one-dimensional solutions of the QHO.

In order to determine the characteristic function of the work distribution of a process, we need to devise the appropriate operation on the light beams and also implement free evolutions. Work performed on the system can be achieved by propagating the light modes through linear optical elements such as lenses, phase masks and spatial light modulators (SLM), which implement unitary transformations. Free evolution can be implemented by an optical transformation called fractional Fourier transform (FRFT). The paraxial optical modes like the Hermite-Guassian and the Laguerre-Gaussian, are eigenfunctions of the FRFT operator. Therefore, under FRFT these modes acquire a global phase that is dependent on their mode labels.

%%%%%%%%%%%%%%%%%%%%%%%%%%%%%%%%%%%%%%%%%%%%%%%%%%%%%%%%%
\begin{figure}[t]
	\includegraphics[width=0.42\textwidth]{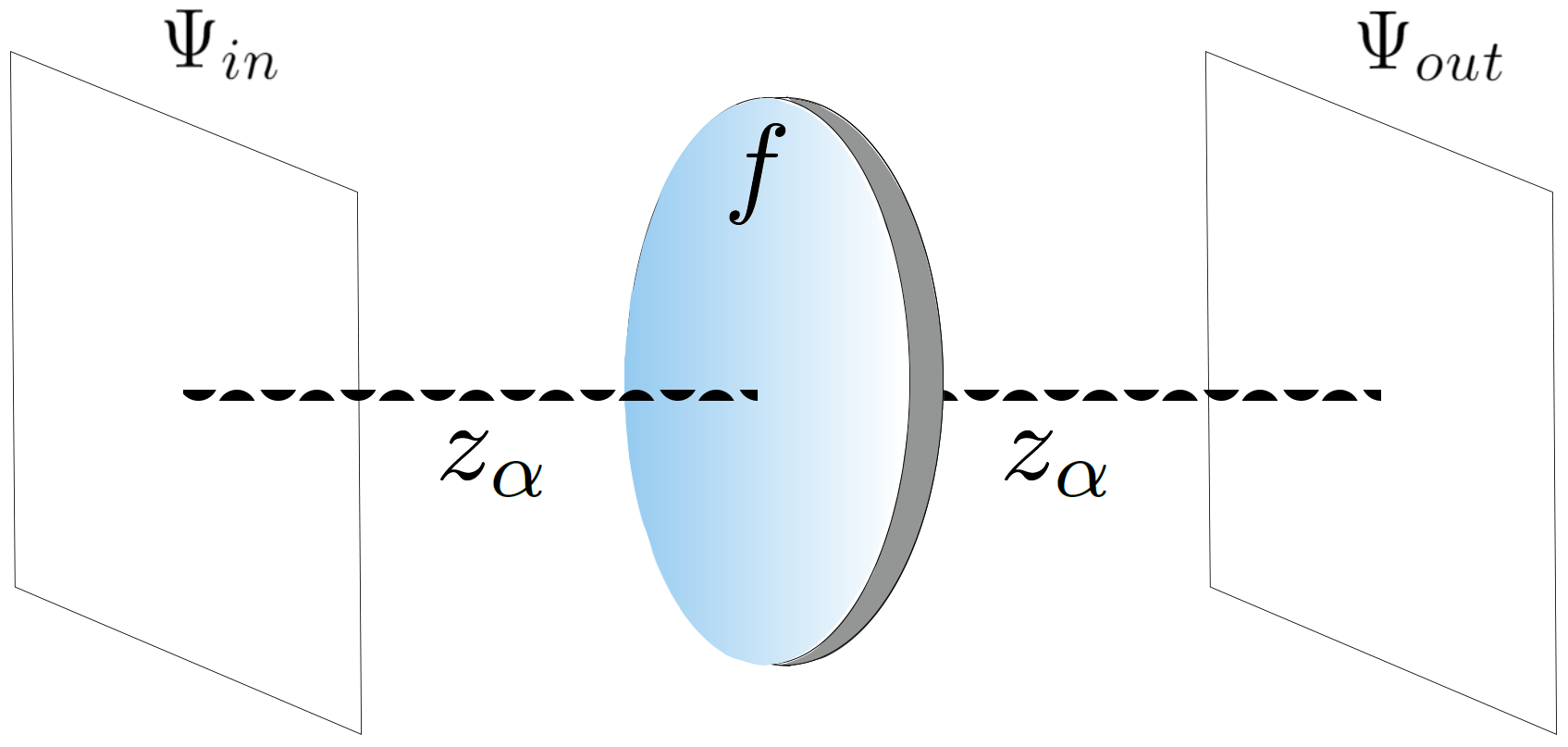}
	\caption{Optical implementation of the fractional Fourier transform. A symmetric lens with focal length $f$ is placed between the input and output plane with a distance $z_{\alpha}$ to each. The field at the output plane is then given by the fractional Fourier transform of the input field.}
	\label{fig:FRFT}
\end{figure}
%%%%%%%%%%%%%%%%%%%%%%%%%%%%%%%%%%%%%%%%%%%%%%%%%%%%%%%%%%

One method for performing the optical FRFT is sketched in Fig. \ref{fig:FRFT}, where the input field is transformed in the output field by free propagation, propagation though a spherical  lens and free propagation again. The FRFT is characterized by a parameter $\alpha \in [0,2\pi]$ that is related to the focal distance $f$ of the lens and the distance $z_{\alpha}$ of each free propagation through 
\begin{eqnarray}
z_{\alpha} = 2 f \sin^2 (\alpha / 2). 
\label{eq:7.2}
\end{eqnarray}
The action of the FRFT can be defined by the operator \cite{Pellat-Finet94}
\begin{eqnarray}
V_{\alpha} = \text{e}^{-i \alpha \frac{\mathbf{P}^2 + \mathbf{X}^2 }{2}}~,
\label{eq:7.3}
\end{eqnarray}
where \textbf{X} and \textbf{P} are the dimensionless position and momentum operators, respectively. This means that the action of the FRFT is a rotation in phase space about the angle $\alpha$ or, equivalently, a free evolution according to the QHO Hamiltonian. The angle $\alpha$ can be controlled by adjusting the distance $z_{\alpha}$ and the focal length $f$. For $\alpha = \pi / 2~ (z_{\alpha}=f)$ this is equal to the optical Fourier transform, a special case of the more general transformation FRFT. Applying the FRFT operator to HG modes, we can verify that they are eigenfunctions of $V_{\alpha}$ \cite{Ozaktas01},
\begin{eqnarray}
V_{\alpha} \phi_{n} = \text{e}^{-i\alpha \varepsilon_n } \phi_n ~.
\label{eq:7.4}
\end{eqnarray}
Here, the modes $\phi_n (x) = \langle x | \phi_n \rangle$ are the $n$-th eigenvector, associated with the $n$-th eigenvalue $\varepsilon_n$, in the position representation. Looking at the action of the FRFT on the light modes we can notice that the transformation encodes the information about the order $n$, and therefore the energy of the system, in the optical phase. This is a key point in the strategy of measuring the characteristic work function using an interferometer.

%%%%%%%%%%%%%%%%%%%%%%%%%%%%%%%%%%%%%%%%%%%%%%%%%%%%%
\begin{figure}[t]
	\includegraphics[width=0.42\textwidth]{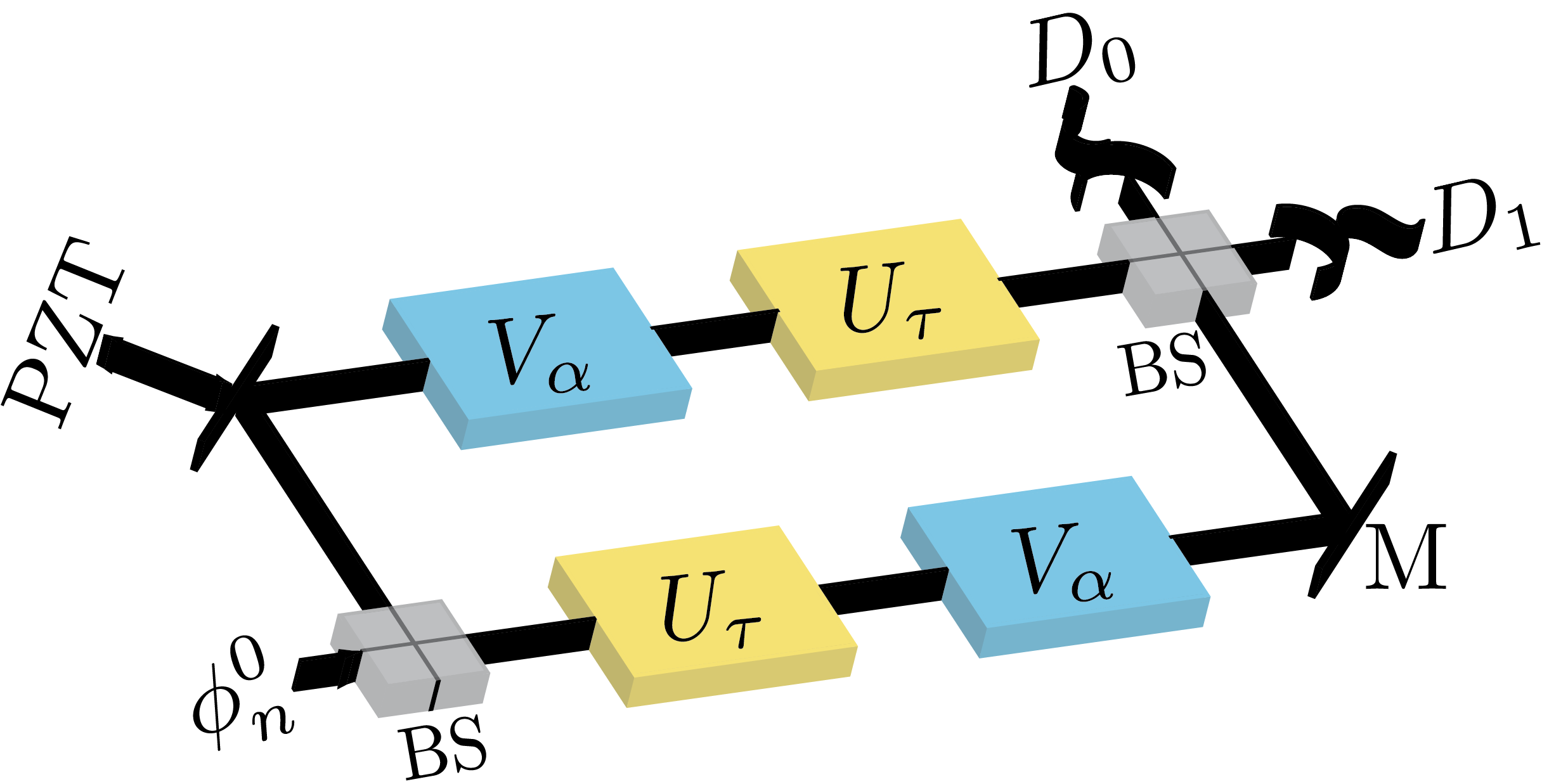}
	\caption{Sketch of the interferometer which implements the protocol to measure the characteristic work function. The input state is split by a beamsplitter (BS) into an upper and a lower path. In the upper path the input state is transformed by the FRFT and then a process $U$ is applied, while in the lower path the order of application is reversed. The output is measured with bulk detectors $\mathcal{D}_1$ and $\mathcal{D}_2$. A PZT is used to control the phase difference to measure the real and imaginary parts. $\mathcal{M}$ is a mirror.}
	\label{fig:interferometer}
\end{figure}
%%%%%%%%%%%%%%%%%%%%%%%%%%%%%%%%%%%%%%%%%%%%%%%%%%%%%%

The sketch of a possible scheme for the interferometer can be seen in Fig. \ref{fig:interferometer}. The input mode $\phi^{0}_{n}$ is prepared in a HG mode. In the upper path of the interferometer, the optical FRFT is applied, corresponding to the QHO free propagation. Afterwards a transformation which is the process acting on the system is performed. After applying these two transformations the state of the upper path can be written as follows
\begin{eqnarray}
\phi^{0}_{n} \rightarrow \phi^{0}_{n} \text{e}^{-i \varepsilon^0_n \alpha} \rightarrow \text{e}^{-i \varepsilon^0_n \alpha} \sum_m c_{m,n} \phi^{\tau}_{m}~.
\label{eq:7.5}
\end{eqnarray}
The expansion coefficients $c_{m,n}$ describe the transition amplitude from the input mode $\phi^0_n$ to the output mode $\phi^{\tau}_m$ after the process and they are defined as
\begin{eqnarray}
c_{m,n} = \displaystyle\int \text{d}x' \text{d}x \left[ \phi^{\tau}_m (x')\right]^* U(x',x,t) \phi^0_n (x)~,
\label{eq:7.6}
\end{eqnarray}
where $U(x',x,t)$ is the coordinate representation of the applied process. The FRFT (in Fig. \ref{fig:interferometer} as $V_{\alpha}$) is implemented like demonstrated in Fig. \ref{fig:FRFT} so that the input plane of the interferometer in the upper path, is transformed onto the plane that is the input plane for the device realizing the process. Another lens can be used to image the output plane of the process onto the output plane of the interferometer, preventing unwanted or uncontrolled evolutions. In this way, the modes evolve in a controlled manner through stroboscopic steps. This imaging adds a constant phase factor to the light modes, which can be controlled by a piezoelectric actuator (PZT) in one of the mirrors so that we won't consider it in the following calculations.

Applying the same treatment to the lower path, but in the reverse order (first the process, then free propagation) we obtain
\begin{eqnarray}
\phi^{0}_{n} \rightarrow \sum_m c_{m,n} \phi^{\tau}_{m} \rightarrow \sum_m c_{m,n} \phi^{\tau}_{m} \text{e}^{-i \varepsilon^{\tau}_m \alpha}~.
\label{eq:7.7}
\end{eqnarray}
Considering 50:50 beam splitters in the input and output of the interferometer, we obtain that the intensity at the output is proportional to \cite{Talarico16}
\begin{eqnarray}
I_n \propto 2 A_n + \Re \left\{ \sum_m |c_{m,n}|^2 \text{e}^{i(\varepsilon^{\tau}_m - \varepsilon^0_n )\alpha} \right\} ~.
\label{eq:7.8}
\end{eqnarray}

Our initial input state shall be prepared in a thermal state as mentioned in Sec. \ref{ssec:work_dist}. This means that the input state is a convex combination of all possible eigenstates with their respective thermal Boltzmann weights. However, because a thermal state is an incoherent mixture of those eigenstates, we can prepare each one of the components of the mixture and apply the process separately. We then sum up over all possible input modes with their weights according to the thermal distribution. In the experiment, a cutoff value can be set for higher order modes, as their probabilities become increasingly small and they do not contribute to the work distribution. Summing up Eq. \eqref{eq:7.8} for all possible input states, we find the intensity proportional to
\begin{eqnarray}
I \propto 2 A + \Re \left\{ \sum_{m,n} p^0_n |c_{m,n}|^2 \text{e}^{i(\varepsilon^{\tau}_m - \varepsilon^0_n )\alpha} \right\}.
\label{eq:7.9}
\end{eqnarray}
If we compare this result to the characteristic work function defined in Eq. \eqref{eq:7.1}, we see that the intensity at the output of the interferometer is, apart from the constant factor $2A$, proportional to its real part, $I \propto 2A + \Re [G(\alpha)]$. We obtain this result considering the upper and lower paths have a zero phase difference. This path difference can be controlled by the PTZ as shown in Fig. \ref{fig:interferometer} and mentioned before. In the same way we can set the path difference to $\pi/2$, which results in the intensity being proportional to the imaginary part of the characteristic function. This means that measuring the intensity at the output of the interferometer we are able to reconstruct the work characteristic function. Calculating the Fourier transform of this function we get the work distribution associated with the considered process. An experimental setup to implement this measurement protocol, as well as detailed calculations, is shown in the Appendix of Ref. \cite{Talarico16}.

%%%%%%%%%%%%%%%%%%%%%%%%%%%%%%%%%%%%%%%%%%%%%%%%%%%%%%%%%%
\subsection{Driven and squeezed harmonic oscillator}
\label{ssec:driven}

As detailed above, some of the transverse modes of a light beam in the paraxial approximation are isomorphic to the energy eigenstates of the 2D quantum harmonic oscillator. We will consider here the one-dimensional driven and squeezed quantum harmonic oscillator. Our calculations will be followed by a suggestion for experimental realization using optical modes and linear optics. The Hamiltonian of the system is
\begin{eqnarray}
\mathbf{H}(t)  = \hbar  \omega \left( \mathbf{a}^{\dagger}\mathbf{a} + \frac{\mathbf{1}}{2}\right)   
&+\eta(t) \mathbf{a}^{\dagger} + \gamma{(t)}\mathbf{a}^{\dagger\,\,2}+ \nonumber\\ 
&+ \eta(t)^{\ast} \mathbf{a} + \gamma^{\ast}{(t)}\mathbf{a}^2,
\label{eq:rel01}
\end{eqnarray}
where $\mathbf{a}^{\dagger}$ and $\mathbf{a}$ are the creation and annihilation operators, which are connected to the quadratures $\hat{x}$ and $\hat{p}$ by
\begin{eqnarray}
\hat{x}=\sqrt{\frac{\hbar}{2m\omega}}\left(\mathbf{a}^{\dagger}+\mathbf{a} \right)
\label{eq:x}
\end{eqnarray}
and
\begin{equation}
\hat{p}=i\sqrt{\frac{\hbar m \omega}{2}}\left(\mathbf{a}^{\dagger}-\mathbf{a} \right).
\label{eq:p1}
\end{equation}
$\omega$ is the oscillator frequency, $\eta{(t)} = |\eta{(t)}|e^{i\Lambda{(t)}}$ is the displacement parameter, and $\gamma{(t)} = |\gamma{(t)}|e^{i\Gamma{(t)}}$ is the squeezing parameter, so that $\Lambda{(t)},\Gamma{(t)} \in \mathbb{R}$. Through the suitable choice of the parameters $\gamma(t)$ and $\eta(t)$ we can control the opening of the potential well and the displacement of the equilibrium point of oscillation. The phases $\Lambda{(t)}$ and $\Gamma{(t)}$ control the orientation in the phase space of the direction in which the displacement and squeezing occur.   These processes are implemented through quenches achieved by linear optical devices.  

In order to explore the work distribution as well as the Jarzynski equality, it is necessary to diagonalize the Hamiltonian (\ref{eq:rel01}). The diagonalized Hamiltonian $\mathbf{H}_{d}(t)$ is connected to $\mathbf{H}(t)$ through a similarity transformation $\mathbf{H}(t)=\mathbf{O}^{\dagger}(t)\mathbf{H}_{d}(t)\mathbf{O}(t)$, where the unitary transformation 
\begin{align}
\mathbf{O}(t) = \mathbf{D}\left[\alpha(t)\right]\mathbf{S}\left[\xi(t)\right] 
\end{align}
is a composition of the displacement operator
\begin{equation*}
\mathbf{D}\left[\alpha(t)\right]=\exp\left[\alpha(t) \mathbf{a}^{\dagger} - \alpha^{*}(t) \mathbf{a}\right]
\nonumber
\end{equation*}
and the squeezing operator
\begin{equation*}
\mathbf{S}\left[\xi(t)\right]=\exp\left\{\frac{r(t)}{2} \left[e^{-i\theta(t)} \mathbf{a}^2-e^{i\theta(t)} \mathbf{a}^{\dagger\,\, 2}\right]\right\}.
\nonumber
\end{equation*}
The displacement parameter is $\alpha(t) = |\alpha(t)|e^{iA(t)}$, with $A(t) \in \mathbb{R}$, and the squeezing parameter $\xi=r(t)e^{i\theta(t)}$ is composed by $r(t) \in \mathbb{R}_{+}$, where its phase is given by $\theta(t) \in \mathbb{R}$. The action of $\mathbf{D}\left[\alpha(t)\right]$ and $\mathbf{S}\left[\xi(t)\right]$ on the creation and annihilation operators is well known \cite{scully1999quantum}
\begin{equation*}
\mathbf{D}^{\dagger}(\alpha)\mathbf{a} \mathbf{D}(\alpha) = \mathbf{a} +\alpha 
\end{equation*}
and
\begin{equation*}
\mathbf{S}(\xi)^{\dagger}\mathbf{a} \mathbf{S}(\xi) = \mathbf{a} \cosh(r) - \mathbf{a}^{\dagger}e^{i\theta}\sinh(r),
\end{equation*}
so that the expression for $\mathbf{a}^{\dagger}$ is attained by Hermitian conjugation of the above formulas. The expression of the diagonalized Hamiltonian is 
\begin{align}
\mathbf{H}_{d}(t) = \hslash\omega{'}(t)\left( \mathbf{a}^{\dagger}\mathbf{a} + \frac{\mathbf{1}}{2}\right)+ \Delta{C(t)},
\label{eq:rel03T}
\end{align}
\noindent
where $\omega{'}(t)=\omega\delta(t)$ is the shifted-frequency, determined by the parameter  \begin{equation}
\delta(t) = \frac{1}{\cosh[2r(t)]},
\label{eq:rel07}
\end{equation}
and $\Delta{C}(t) =  - \hslash\omega|\alpha(t)|^{2}$ is a shift on the energy. Actually, to obtain the final form of $\mathbf{H}_{d}(t)=\mathbf{O}(t)\mathbf{H}(t)\mathbf{O}^{\dagger}(t)$ we imposed that, after the transformation $\mathbf{O}(t)$ on $\mathbf{H}(t)$, the coefficients multiplying the operators $\mathbf{a},\mathbf{a}^{\dagger}, \mathbf{a}^2$, and $\left( \mathbf{a}^{\dagger}\right)^{2}$ are null. Therefore, the connection between the parameters of the unitary transformations and the parameters of the Hamiltonian (\ref{eq:rel01}) is determined by the relations
\begin{align}
|\alpha|&=\frac{|\eta|}{\omega^{2} - 4|\gamma|^{2}}\sqrt{4|\gamma|^{2} +  \omega^{2} - 4|\gamma|\omega\cos{(\Lambda - 2\Gamma)}},\\
A&= \arctan \left[\frac{2|\gamma|\sin(\Lambda - \Gamma) - \omega\sin(\Lambda)}{2|\gamma|\cos(\Lambda - \Gamma) - \omega\cos(\Lambda)}\right],\\
\theta&=\Gamma,\\
r&=  \frac{1}{2}\arctanh{\left(\frac{2|\gamma|}{\omega}\right)} \label{eq:r}.
\end{align}
Naturally, the diagonalization process imposes restrictions on the degree of squeezing $|\gamma| \in [0, \sfrac{\omega}{2})$, which in turn, changes the shifted-frequency $\omega{'}(t)$ through the parameter $\delta(t)$ [see Eqs. (\ref{eq:rel07}) and (\ref{eq:r})].

%%%%%%%%%%%%%%%%%%%%%%%%%%%%%%%%%%%%%%%%%%%%%%%%%%%%%%%%%%%%%%%%
\begin{figure}[bh]
	\centering
	\includegraphics[width=1.0\linewidth]{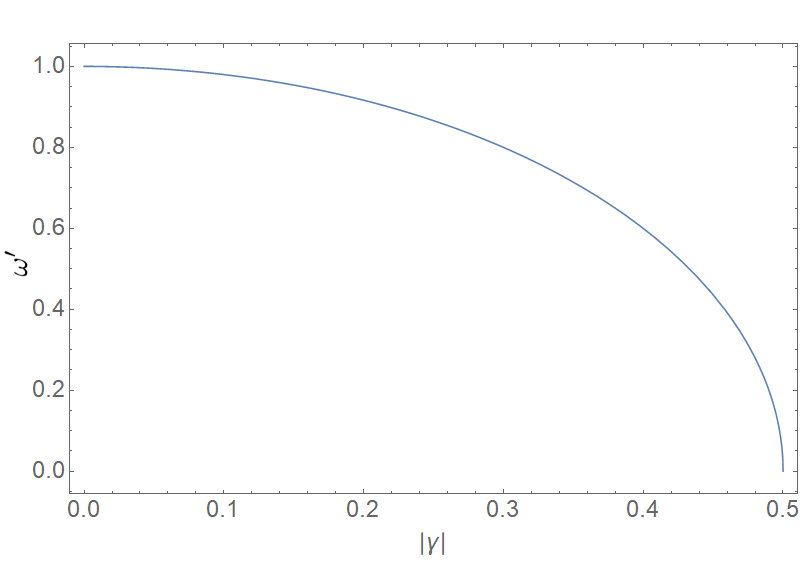}
	\caption{The intensity of the shifted-frequency $\omega{'} = \omega\delta$ as function of the squeezing parameter $|\gamma|$ written in units of $\omega$.} 
	\label{fig:frel01}
\end{figure}
%%%%%%%%%%%%%%%%%%%%%%%%%%%%%%%%%%%%%%%%%%%%%%%%%%%%%%%%%%%%%%%%

\begin{figure}[ht]
	\centering
	\includegraphics[width=1.0\linewidth]{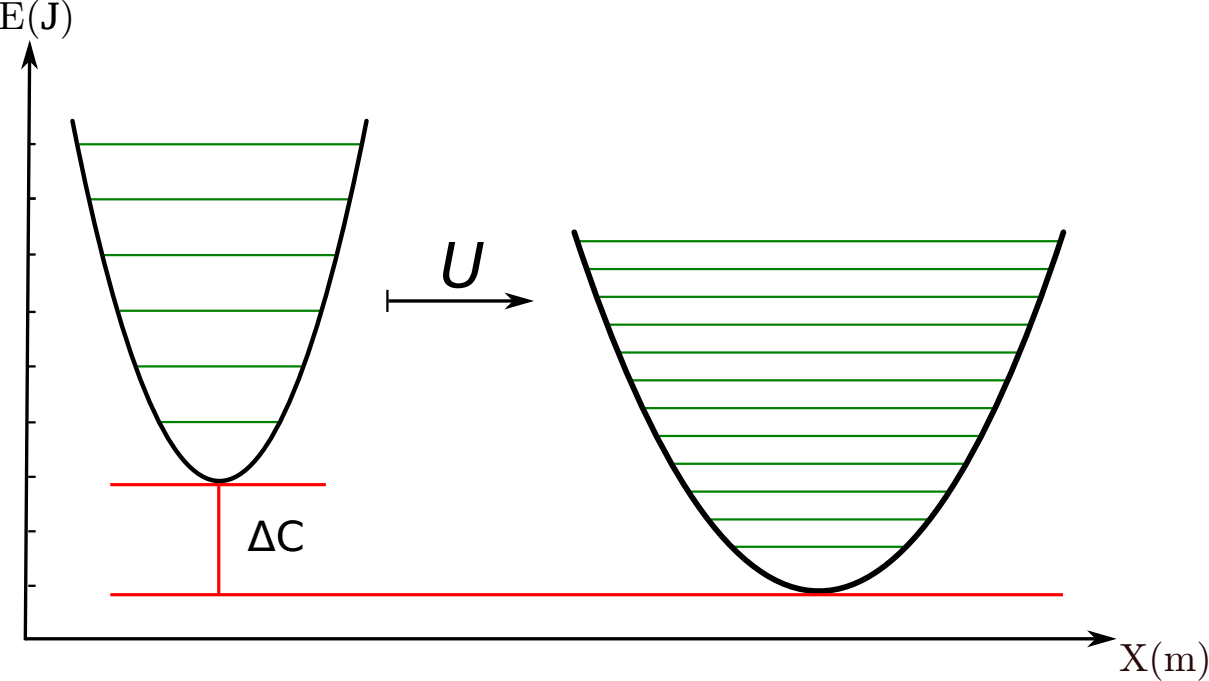} %[scale=0.36]
	\caption{Change in the spectrum of the harmonic oscillator due to the sudden quench in the Hamiltonian promoted by the displacement and squeezing terms in the Hamiltonian (\ref{eq:rel01}). Effectively, by increasing the value of $|\gamma|$ the system becomes decompressed, as can be seen by the change in $\omega{'}$.} 
	\label{fig:frel02}
\end{figure}

In Fig. \ref{fig:frel01} it is shown the dependence of $\omega{'}(t)$ on the squeezing parameter $|\gamma|$, where we notice that increasing the degree of squeezing the new frequency becomes smaller. Such effect is better visualized in Fig. \ref{fig:frel02}, where the spectrum of the driven and squeezed harmonic oscillator
\begin{equation}
E_n = \hslash\omega{'}\left(n+\frac{1}{2}\right)+ \Delta{C},\hspace{0.5cm} n=0,1,2,\dots
\end{equation}
is plotted for $\eta=0$ and $\gamma=0$ (\textit{left}) and for $\eta \ne 0$ and $\gamma \ne 0$ (\textit{right}). We observe that as $|\gamma|$ increases the potential well becomes wider, or equivalently, the effective frequency of oscillation becomes smaller. The opposite effect, in which the potential well becomes tighter, can be achieved if the initial and final configurations are interchanged, i.e., they satisfy $|\gamma(0)| > |\gamma(\tau)|$.

%%%%%%%%%%%%%%%%%%%%%%%%%%%%%%%%%%%%%%%%%%%%%%%%%%%%%%%%%%%%%%%%%%%
%%%%%%%%%%%%%%%%%%%%%%%%%%%%%%%%%%%%%%%%%%%%%%%%%%%%%%%%%%%%%%%%%%%

As we are considering a sudden change of the Hamiltonian parameters, the unitary evolution operator is $\mathbf{U}_{\tau} \backsimeq \mathds{1}$, so that the conditional probabilities given by Eq. (\ref{eq:prob-work}) can be written as
\begin{align}
p_{m|n} = |\langle \varepsilon_{m}^{\tau}|\varepsilon_{n}^{0}\rangle|^2,
\label{eq:rel15}
\end{align}
where the eigenstates of the initial Hamiltonian are $|\varepsilon^{0}_{n}\rangle = S^{\dagger}\left[\xi(0)\right]D^{\dagger}\left[\alpha(0)\right]|n\rangle$ and of the final Hamiltonian are $|\varepsilon^{\tau}_{m}\rangle = S^{\dagger}\left[\xi(\tau)\right]D^{\dagger}\left[\alpha(\tau)\right]|m\rangle$, with $|n\rangle$($|m\rangle$) being the Fock state. Rewriting the conditional probability (\ref{eq:rel15}) as function of the displacement and squeezing operators, we have
\begin{equation}
p_{m|n} = |\langle m|D\left[\alpha(\tau)\right]S\left[\xi(\tau)\right]S^{\dagger}\left[\xi(0)\right]D^{\dagger}\left[\alpha(0)\right]|n\rangle|^2.
\label{eq:rel31}
\end{equation}

Now we have all the necessary quantities in order to analyze the work distribution in our system, however, to verify Jazynski equality, we need to evaluate the Helmholtz free energy, defined by
\begin{equation}
\Delta{F} = \frac{1}{\beta}\ln\left[ \frac{Z(\mathbf{H}(0))}{Z(\mathbf{H}(\tau))} \right],
\label{eq:rel33a}
\end{equation}
which, in our case, takes the form 
\begin{equation}
\Delta{F}=  \frac{\hslash\omega}{2}\left(\delta(\tau)-\delta(0)\right) + \ln \left(\frac{1 -e^{-\beta\hslash\omega\delta(\tau)}}{1 -e^{-\beta\hslash\omega\delta(0)}}\right)+\Delta{C}.
\label{eq:rel34}
\end{equation}

%%%%%%%%%%%%%%%%%%%%%%%%%%%%%%%%%%%%%%%%%%%%%%%%%%%%%%%%%%%%%%%%%%
\subsubsection{Displacement effects}

\begin{figure}[h]
	\includegraphics[width=0.4\textwidth]{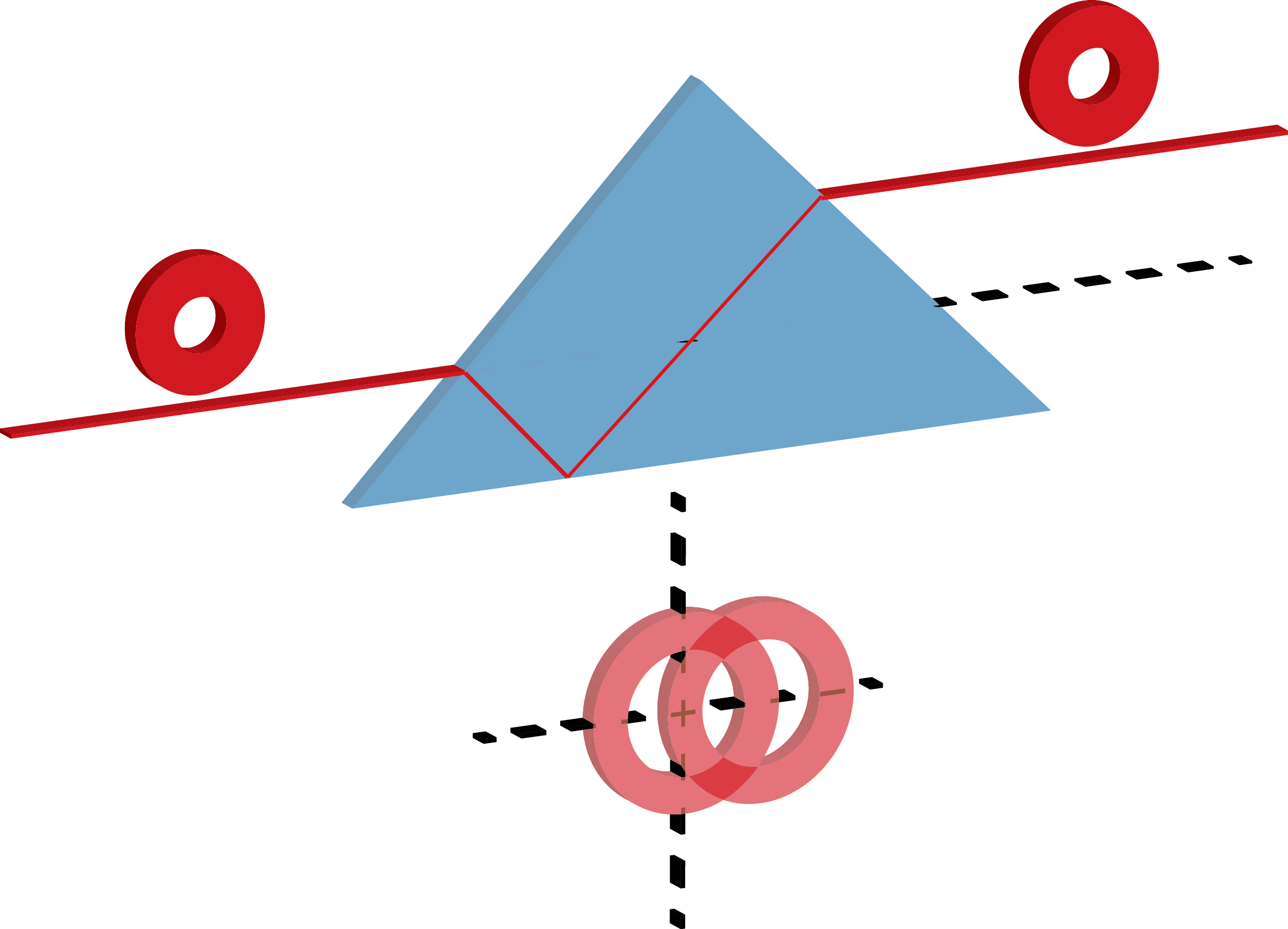}
	\caption{In the upper part, the light beam is displaced by internal reflection inside a prism. In the lower part, the overlap between displaced modes is illustrated.}
	\label{fig:osc_displac}
\end{figure}

\begin{figure}[ht]
	\centering
	\includegraphics[width=\linewidth]{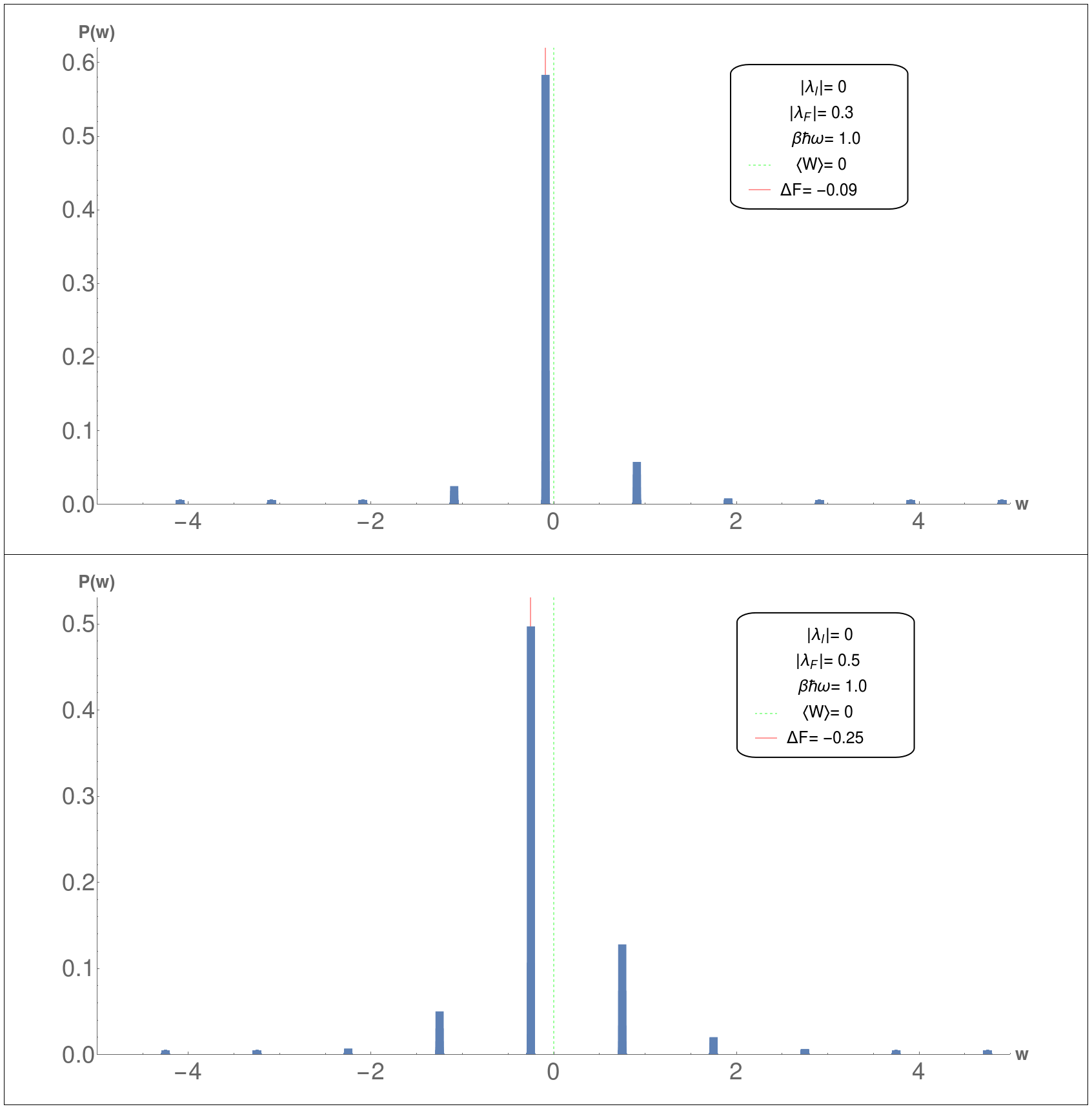}
	\caption{Work distribution for the driven harmonic oscillator with $\gamma(t)=0$, $\Lambda(t)=0$ for $\forall t$, $\eta(0)=0$, $\beta\hbar\omega=1$, and $\eta(\tau)=0.3\omega$ for the upper figure, while $\eta(\tau)=0.5\omega$ for the lower figure. The red solid vertical line marks the free energy value and the green dotted vertical line marks the average work performed on the system.} 
	\label{fig:displacmente01}
\end{figure}

Let us first consider the particular case in which $\gamma(t)=0$ in Hamiltonian (\ref{eq:rel01}). In this situation the diagonalization is attained through the displacement operator only, that is why we named such contributions \textit{displacement effects}. In Ref. \cite{Talarico16} it was reported the work distribution when the linear momentum of the oscillator is displaced by a constant value $p\rightarrow p+p_0$ with $\Delta F=0$. Here we analyze two cases in Fig. \ref{fig:displacmente01} where initially $\eta(0)=0$ and after the sudden quench $\eta(\tau)=0.3\omega$ (top figure) and $\eta(\tau)=0.5\omega$ (bottom figure). As the phase of $\eta(t)$ is null, $\Lambda (t)=0$, the average position of the oscillator is displaced by $x_0=\eta\sqrt{\frac{2\hbar}{m\omega^3}}$. These processes imprint a negative free energy variation $\Delta F=-\frac{\hbar \eta^2}{\omega}$ and the average work performed on the system is null. In order to illustrate that, we devise in Fig. \ref{fig:osc_displac} an experimental setup for realizing the displacement operations with paraxial light modes representing energy eigenstates of the two-dimensional harmonic oscillator. The system here is the wave-front of the light beam, which has its direction of propagation changed by the prism. This process is realized by propagating a light beam prepared in one of the Laguerre-Gaussian modes through a prism that displaces its axis parallel to the incident one. This simple operation changes the Hamiltonian, as the origin of the coordinate system is changed. The orbital angular momentum, which gives the information about the energy, depends on the origin and there will be coupling between the displaced and the non-displaced family of modes.

%%%%%%%%%%%%%%%%%%%%%%%%%%%%%%%%%%%%%%%%%%%%%%%%%%%%%%%%%%%%%%%%%%%
%%%%%%%%%%%%%%%%%%%%%%%%%%%%%%%%%%%%%%%%%%%%%%%%%%%%%%%%%%%%%%%%%%

Another important aspect to be analyzed in the sudden displacement case is the work distribution. We observe in Fig. \ref{fig:displacmente01} that the effect of the driven term is to broaden the work distribution and to displace the position of the center of the distribution. This result is expected since higher energy modes of the harmonic oscillator are excited after the application of the linear quench enabling new transitions among the energy eigenstates. As such displacement affects all energy eigenstates, we expect that the center of the distribution is displaced proportionally to the intensity of $\eta(\tau)$.

%%%%%%%%%%%%%%%%%%%%%%%%%%%%%%%%%%%%%%%%%%%%%%%%%%%%%%%%%%%%%%%%%%    
%%%%%%%%%%%%%%%%%%%%%%%%%%%%%%%%%%%%%%%%%%%%%%%%%%%%%%%%%%%%%%%%%

%%%%%%%%%%%%%%%%%%%%%%%%%%%%%%%%%%
\subsubsection{Squeezing effects}

We now analyze how sudden squeezing processes affect the work distribution. In this situation we turned off the driven parameter $\eta(t)=0$ in Hamiltonian (\ref{eq:rel01}) and consider two particular scenarios. In the upper part of Fig. (\ref{fig:squeezing}) we set $\gamma(0)=0$ and $\gamma(\tau)=0.3\omega$ for $\beta\hbar\omega=0.5$. This sudden quench describes a decompression situation in which the potential well of the harmonic oscillator is opened with respect to the initial condition, $\omega \rightarrow \omega^{'}=0.8\omega$. The average work performed on the system in this case is null while the variation of free energy is negative. In the scenario in which the opposite process occurs, as shown in the lower part of Fig. (\ref{fig:squeezing}), we set $\gamma(0)=0.3\omega$ and $\gamma(\tau)=0$ for $\beta\hbar\omega=0.5$. Now, the potential well becomes narrower and consequently the effective frequency increases after the quench, $\omega^{'}=0.8\omega \rightarrow \omega$. The average work performed on the system and the free energy variation are positive, which reflects the work performed on the light beam by the optical devices. 

\begin{figure}[!h]
	\includegraphics[width=0.42\textwidth]{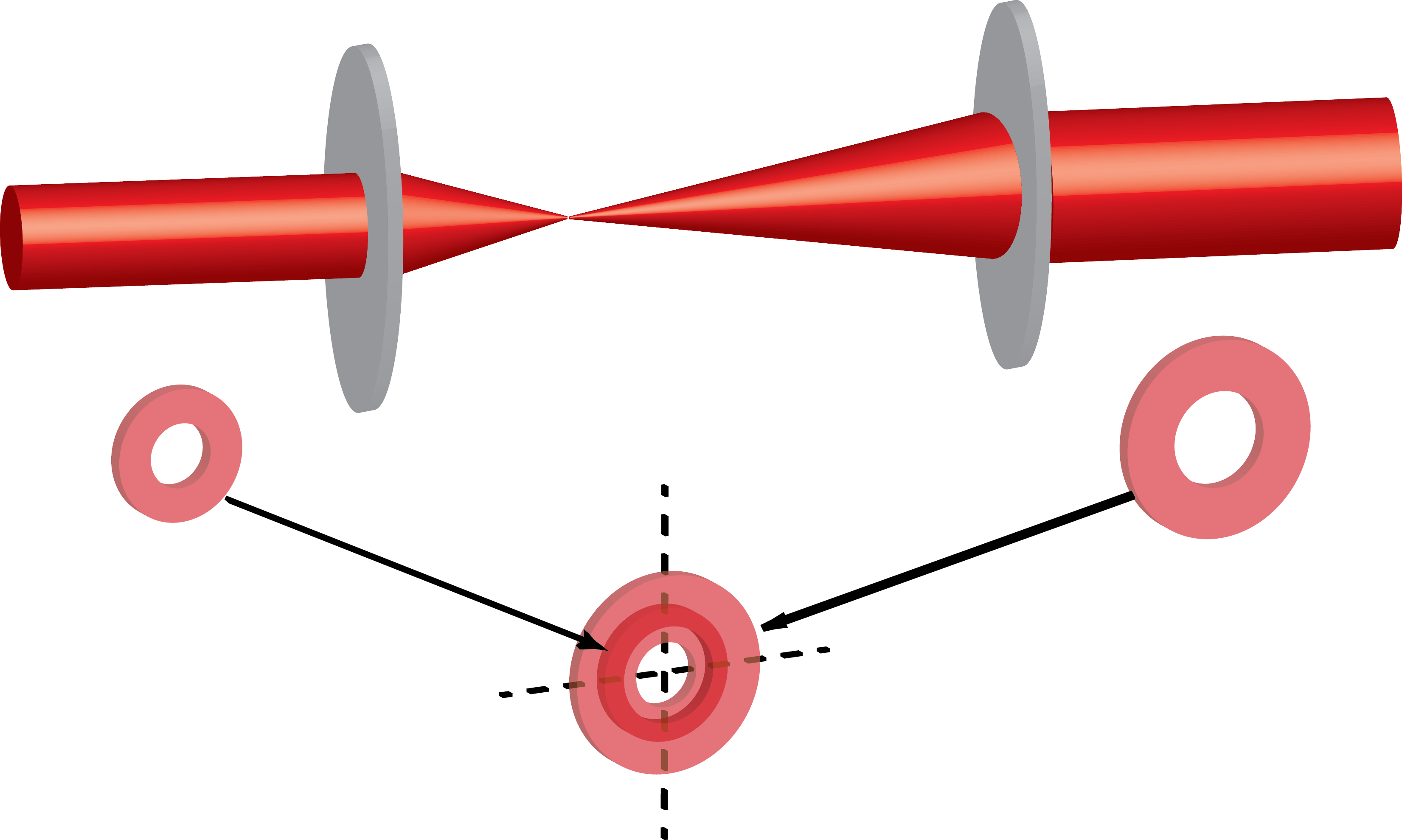}
	\caption{\label{fig:osc_squeez}
		The light beam is expanded or compressed with a telescope. In the lower part, the overlap between smaller and bigger modes is illustrated.}
\end{figure}

\begin{figure}[ht]
	\centering
	\includegraphics[width=0.9\linewidth]{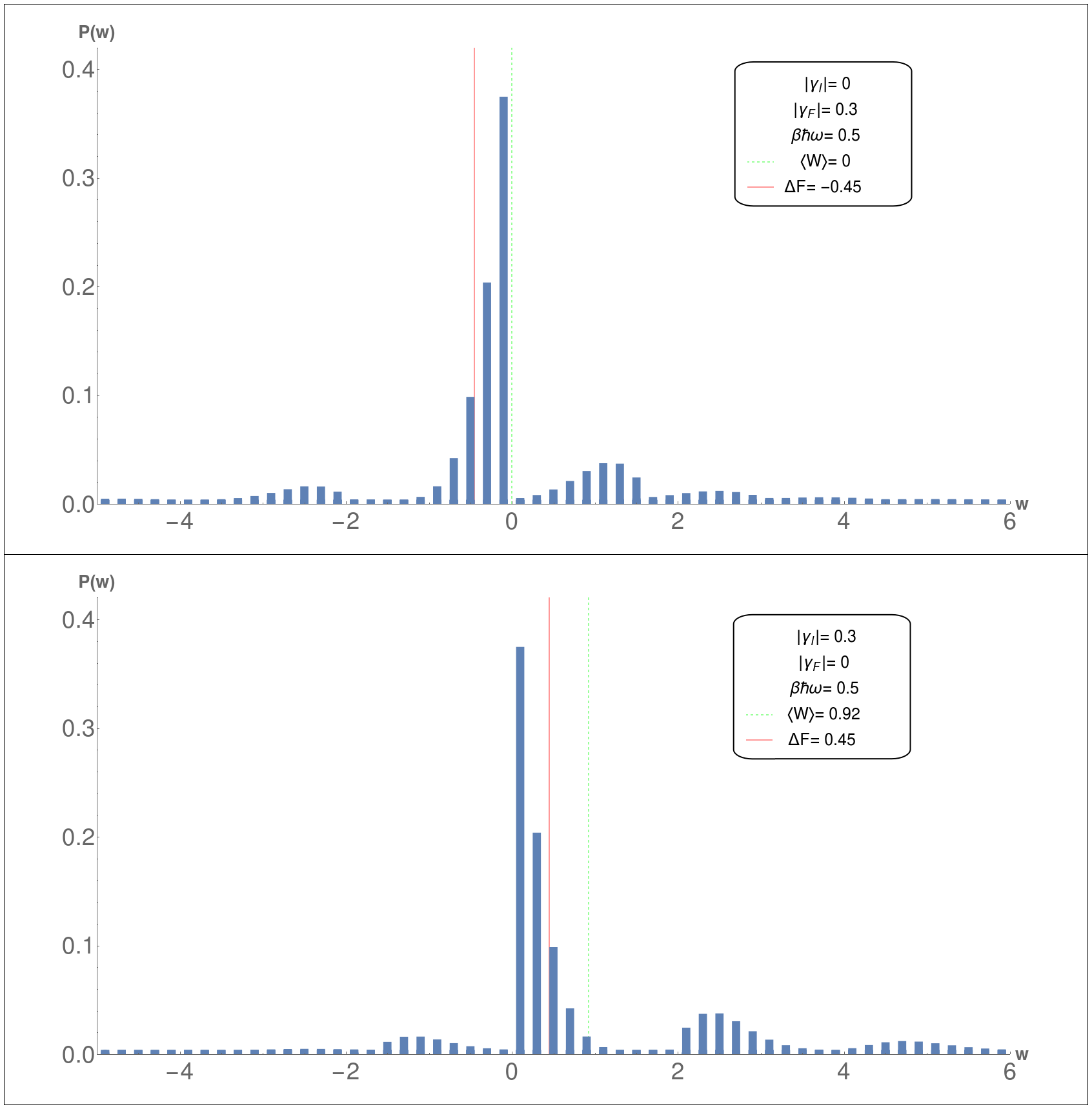}
	\caption{Work distribution of the harmonic oscillator after sudden decompressed and squeezing quenches. The parameters are set to $\beta\hbar\omega=0.5$ and $\eta(t)=\Lambda(t)=\Gamma(t)=0$ for $\forall t \in [0,\tau]$ for both figures. The upper figure shows a sudden decompressed quench with $\gamma(0)=0$ and $\gamma(\tau)=0.3\omega$, so that the characteristic frequency of the oscillator after the quench is $\omega_F=0.8\omega_0$. The lower figure shows the opposite case, i.e., a squeezing process in which $\gamma(0)=0.3\omega$ and $\gamma(\tau)=0$. The red solid and green dotted vertical lines point the values of Helmholtz free energy and the average work.} 
	\label{fig:squeezing}
\end{figure}

Fig. \ref{fig:osc_squeez} shows the sketch of the setup for realizing squeezing or decompression operations. As before, the input beam is prepared in one of the Laguerre-Gaussian modes and sent through a beam expander (left) or compressor (right). This will change the Hamiltonian, as the expanded (compressed) beams represent a new family of modes with larger (smaller) beam waist. Therefore, there will be overlap between one single mode of one family and some modes in the other family.

%%%%%%%%%%%%%%%%%%%%%%%%%%%%%%%%%%%%%%%%%%%%%%%%%%%%%%%%%%%%%%%%%%%
%%%%%%%%%%%%%%%%%%%%%%%%%%%%%%%%%%%%%%%%%%%%%%%%%%%%%%%%%%%%%%%%%%%

As we can see in Fig. \ref{fig:squeezing}, the profile of the work distribution of the quantum system after the sudden squeezing or decompression quenches are quite different from the displacement quenches. We observed that some degree of asymmetry in the distribution is introduced depending on the value of $\gamma(t)$ at the beginning and at the end of the process. Moreover the work distribution present oscillations and revivals. Another feature described in Fig. \ref{fig:squeezing} is that the squeezing quench (lower part) increases the chance of observing a violation the second law of thermodynamics (in a single run) with higher probability than the decompressed quench (upper part).

%%%%%%%%%%%%%%%%%%%%%%%%%%%%%%%%%%%%%%%%%%%%%%%%%%%%%%%%%%%%%%%%%%
%%%%%%%%%%%%%%%%%%%%%%%%%%%%%%%%%%%%%%%%%%%%%%%%%%%%%%%%%%%%%%%%%

In table \ref{tab:jarzynski} we check the validity of the results presented in Figs. \ref{fig:displacmente01} and \ref{fig:squeezing} through the numerical verification of the Jarzynski equality and the normalization of the work distribution.

\begingroup\makeatletter\def\f@size{8}\check@mathfonts
\begin{table}[h]
\centering
\begin{tabular}{|cccccccc|}\hline
$\beta\hslash\omega$ & $\lambda(t_F)$&$\gamma(t_I)$&$\gamma(t_F)$&$\langle \mathcal{W} \rangle $&$\Delta{F}$& $\langle e^{-\beta(\mathcal{W} - \Delta{F})}\rangle$&||$P(\mathcal{W})$|  \\ \hline\hline
 1.0&0.3&0&0&0&-0.09&1.00&1.00\\ \hline
 1.0&0.5&0&0&0&-0.25&1.00&1.00\\ \hline
 0.5&0&0&0.3&0&-0.45&0.99&1.00\\ \hline
 0.5&0&0.3&0&0.92&0.45&1.00&1.00\\ \hline
 \end{tabular}
\caption{Numerical results for the Helmholtz free energy, average work, Jarzynski equality, and the normalization of the work probability distribution for the set of parameters used in Figs. \ref{fig:displacmente01} and \ref{fig:squeezing}. In all cases $\lambda(t_I)=0$.}
\label{tab:jarzynski}
\end{table}
\endgroup

%%%%%%%%%%%%%%%%%%%%%%%%%%%%%%%%%%%%%%%%%%%%%%%%%%%%%%%%%%%%%
\section{Experimental studies}
\label{sec:exp}

In this section we provide some examples illustrating how an all-optical setup can be helpful in the investigation of thermodynamics at the quantum level. 

%%%%%%%%%%%%%%%%%%%%%%%%%%%%%%%%%%%%%%%%%%%%%%%%%%%%%%%%%%%%%
\subsection{Simulation of single-qubit thermometry}
\label{ssec:qubitthermo}

The first one \cite{Barbieri17} analyzes the use of a single qubit as a thermometer. In standard thermodynamics, the temperature is defined only for systems in equilibrium with its surroundings acting as a thermal bath. A usual method to measure the temperature is to a thermometer, which does not affect the equilibrium conditions. However, when the system becomes smaller, the thermometer needs to be even smaller in comparison to the thermal bath.

In order to reduce the scale of the thermometers, Jevtic \textit{et al.} \cite{jevtic15} proposed a model where you can use a single qubit to obtain the information about two temperatures of a bosonic bath. Mancino \textit{et al.} \cite{Barbieri17} presented an experimental investigation of this model using a laser beam, interferometers, and photodiodes. They implemented and measured optical thermal states prepared in the polarization degree of freedom.

They employed a linear-optical-simulator to emulate the interaction between one qubit and a thermal bath. When the qubit is isolated from the bath, $\ket{0}$ is the excited state and $\ket{1}$ is the ground state, and $\hbar\omega$ is the  energy difference between levels.

To simulate the qubit in the presence of a thermal bath, they used a quantum channel to implement the excitation (decay) of the ground (excited) state. The process that can be applied to realize this interaction is the generalized amplitude damping channel (GAD), described by two pairs of Kraus operators. The first one is $E_{0}$ and $E_{1}$, which describe the standard decay of the excited state of the qubit into a "cold" bath $T_{1}$. The second pair, $E_{2}$ and $E_{3}$, is related to the inverse process, where the qubit is placed in a "hot" bath $T_{2}$ and goes from the ground to the excited state, and $T_{2}>T_{1}$. With the Kraus operators it is possible to reconstruct the process.

\begin{figure}[!h]
\centering
\includegraphics[width=0.42\textwidth]{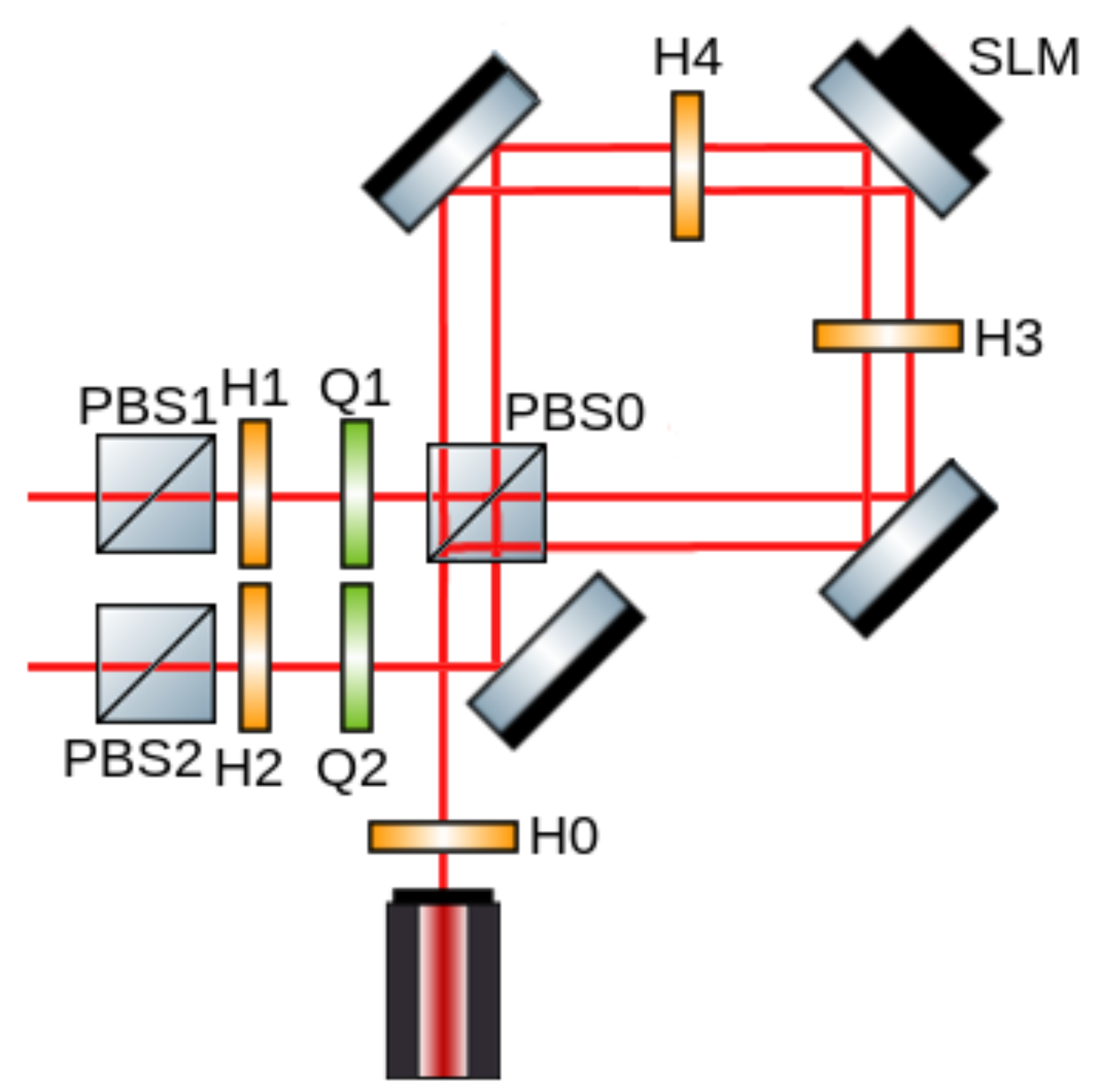}]
\caption{Experimental setup for implementing a generalized amplitude damping channel.}
\label{fig:setup_thermometry}
\end{figure}

The setup is sketched in Fig. \ref{fig:setup_thermometry}. It consists of a Sagnac-like interferometer, which implements the GAD. One of the mirrors of the Sagnac is replaced with a spatial light modulator (SLM). The ground (excited) state, by convention, the state $\ket{1}$ ($\ket{0}$) is the vertical $\ket{V}$ (horizontal $\ket{H}$) polarization state.

%%%%%%%%%%%%%%%%%%%%%%%%%%%%%%%%%%%%%%%%%%%%%%%%%%%%%%%
%%%%%%%%%%%%%%%%%%%%%%%%%%%%%%%%%%%%%%%%%%%%%%%%%%%%%%

Three different input states where prepared for the thermometer: (i) the ground state of the qubit ($\ket{V}$) represent the situation where the thermometer will be heated up by the hot bath; (ii) the excited state ($\ket{H}$) represents the situation where the thermometer will be cooled down by the cold bath; (iii) The third state ($\ket{+}$) is a superposition between hot and cold thermometer state. Each input state is sent through the process, which is the cold or the hot bath. The goal is to analyze the output state of the polarization qubit using state tomography, and from the population difference determine if the bath was hot or cold. The interaction time is simulated by changing the parameter $p$ in the channels that are implementing the baths. 

The results show that, for short interaction times, the discrimination is optimal and approaches the theoretical prediction. They also show that all three input states are equally suited to the task and the coherent superposition input state presents no gain in this scenario. In conclusion, the polarization qubit can be used as a thermometer for these emulated thermal baths.

%%%%%%%%%%%%%%%%%%%%%%%%%%%%%%%%%%%%%%%%%%%%%%%%%%%%%%%%%%%%
\subsection{Photonic Maxwell's demon}
\label{ssec:maxdem}

\begin{figure}[!h]
\centering
\includegraphics[width=0.5\textwidth]{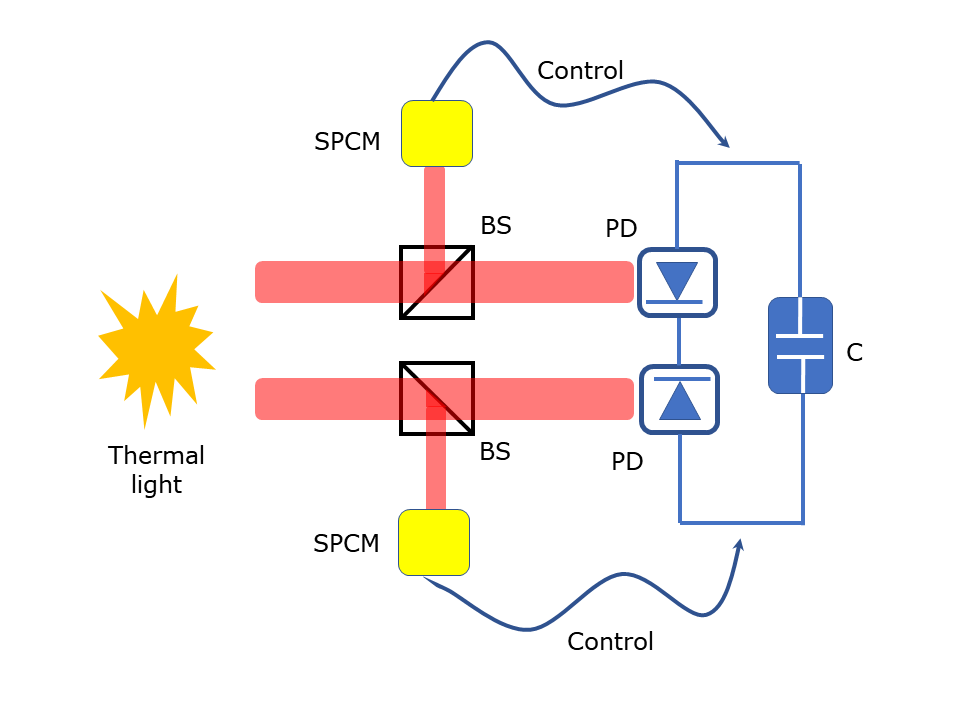}]
\caption{\label{fig:setup_maxwell_demon}
Schematic representation of the experimental set-up for demonstrating a photonic Maxwell's demon.}
\end{figure}

Let us now discuss one more experiment employing an optical set-up. This time, instead of polarization, the authors use the photon number or energy degree of freedom to experimentally address a paradigmatic problem in thermodynamics, Maxwell's demon. Mihai \textit{et al} \cite{Barbieri16}, used the setup schematically shown in fig. \ref{fig:setup_maxwell_demon} to demonstrate that it is possible to extract work from an intense pseudo-thermal light source and use it to charge a capacitor. 
 
%%%%%%%%%%%%%%%%%%%%%%%%%%%%%%%%%%%%%%%%%%%%%%%%%%%%%%%%%%%% 
%%%%%%%%%%%%%%%%%%%%%%%%%%%%%%%%%%%%%%%%%%%%%%%%%%%%%%%%%%%%%

The experimental set-up is illustrated in Fig. \ref{fig:setup_maxwell_demon}. Light coming from the pseudo-thermal source is split into two beams with the same average intensities. Each one is directed to a high transmission beam splitter (BS). The transmitted light is detected by a photodiode (PD) and the reflected light is detected by a single-photon counting module (SPCM). The PD converts light into electric current. Without accounting for the information coming from the SPCM, the average voltage across the capacitor (C) is zero. However, the event of a photon count in the SPCM is correlated with the intensity fluctuation of the transmitted beam, and this can be used to switch the capacitors polarity according to the conditional counts in the SPCM.

As the thermal light has the photon bunching effect, when a photon is reflected from the BS into the SPCM, there is a higher probability that the number of photons in the transmitted beam fluctuates above the average. Therefore, using the information coming from both SPCMs and feed forward, the polarity of the capacitor can be properly switched. One count in one of the SPCMs and no count in the other helps in charging the capacitor, while two counts or two no-counts do not contribute to the charging process. The authors provided a proof-of-principle by measuring the intensity difference between the PDs conditioned on the count and no-count events in the SPCMs.They do not actually implement the feed-forward control to demonstrate effective charging of the capacitor. 

%%%%%%%%%%%%%%%%%%%%%%%%%%%%%%%%%%%%%%%%%%%%%%%%%%%%%%%%%%%%
\subsection{Work distribution with paraxial light modes: Two-point measurement protocol}
\label{ssec:twomeasure}

Another experimental method to measure the work distribution of a process acting on a system simulated with paraxial light modes is through projective measurements, by directly considering the two-point-measurements protocol presented in Sec. \ref{ssec:work_dist}. In the previous section, we introduced a method that recovers the information about the energy of the system through interference in order to avoid direct energy measurements. Here, we want to show how to reconstruct the work distribution by measuring the energy levels of the system through projective measurements after a process has been applied. The system and the applied process can be simulated experimentally with paraxial beams due to the analogy between the paraxial equation and the Schr\"{o}dinger equation (see Sec. \ref{sec:analogy}). A family of light modes that are solutions to the paraxial wave equation and, therefore, simulating a quantum harmonical oscillator are the Laguerre-Gaussian (LG) modes. Their energy eigenvalues are \cite{Allen92}
\begin{eqnarray}
\varepsilon_{\ell p} = ( \abs{\ell} +2p +1 ) \hbar \omega~,
\end{eqnarray}
where $\ell$ and $p$ are the azimuthal and radial quantum numbers, respectively, which correspond to the azimuthal and radial indexes to identify the elements of the LG basis of modes. If we consider only modes with $p=0$, their eigenvalues reduce to
\begin{eqnarray}
\varepsilon_{\ell} = ( \abs{\ell} + 1 )\hbar \omega
\end{eqnarray}
and therefore depend on $\ell$ only. LG light modes contain a orbital angular momentum (OAM) \cite{Padgett00}. The amount of OAM per photon of each mode is determined by the quantum number $\ell$. This means that a projection onto the OAM basis is equivalent to a projection onto the energy eigenbasis. If we restrict ourselves to processes that only change $\ell$ and then project the final state in the OAM basis, the work done on the system in the transition from an initial $\ell$ to a final $\ell '$ can be defined as
\begin{eqnarray}
W_{\ell \ell '} = ( \abs{\ell '} - \abs{\ell} ) \hbar \omega~.
\end{eqnarray}
Using this definition we can calculate the work distribution in Eq. \eqref{eq:rel11}. The probabilities $p_{m,n}$ of that distribution are in this analogy the probabilities $p_{\ell , \ell'} = p_{\ell} p_{\ell' | \ell}$. 
This is the probability to observe the transition $\ell \rightarrow \ell '$, $p_{\ell}$ is the probability of having $\ell$ as an input and $p_{\ell'| \ell}$ the probability of observing $\ell'$ at the output given that $\ell$ was the input. 
Those probabilities $p_{\ell}$ are the thermal probabilities defined in equation \eqref{eq:thermal}, where the denominator $Z_0$ is the partition function. The partition function is defined in terms of a sum over all possible energy states. In the analogy between OAM states and QHO energy eigenstates we have to consider that there are OAM states with negative values of $\ell$, but for a HO there are no negative energies. Those states with negative $\ell$ have the same energy and the same probabilities as their positive equivalent, $p_{\ell}=p_{-\ell}$. This results in a degeneracy 2 for all energy states, except for $\ell = 0$ and therefore in a different partition function to adjust the thermal probabilities of each mode. We obtain for the partition function \cite{Araujo18}
\begin{eqnarray}
Z_0 = \sum_{\ell} \text{e}^{-\beta \varepsilon_\ell^{0}} = \left( \text{e}^{\beta \hbar \omega} \tanh \frac{\beta \hbar \omega}{2} \right)~.
\end{eqnarray}

The first step of the two-point-measurement protocol is to prepare the initial state, which is a thermal one in the OAM basis. In an optical setup, this can be done by directing a Gaussian laser beam onto a SLM, which modulates the light beam with a programmed phase mask to generate a LG mode with the desired OAM. The input modes don't need to be produced with their respective thermal probabilities because a thermal state is a incoherent mixtures of all possible input states (see previous section). We can generate every one of the basis state and apply a process independently to each and multiply by its thermal probability afterwards in order to calculate the work distribution. A range of input states like $-10 \leqslant \ell \leqslant 10$ can be chosen because higher order states contribute to the work distribution negligibly and their Boltzmann weights can be ignored.

In the second step of the protocol, a process acting on the system is applied. This can be done by a second SLM, using another phase mask that introduces transitions between different energy levels and thus, perform work on the system.

\begin{figure}
\includegraphics[width=0.42\textwidth]{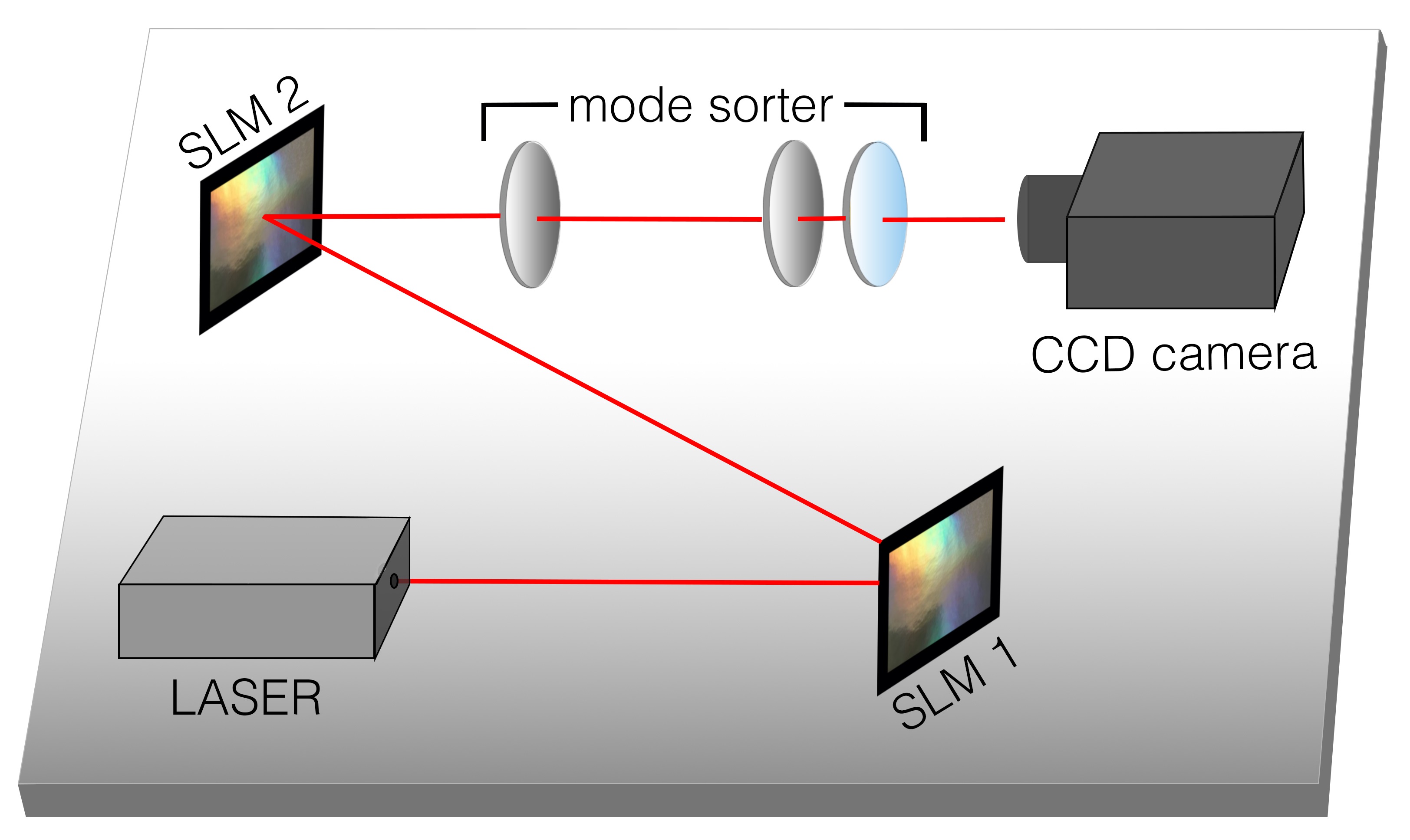}
\caption{Experimental setup for an optical implementation of the two-point measurement protocol to obtain the work distribution. A laser emitting a Gaussian mode is directed to SLM1 which generates the LG input modes. SLM2 applies a process that introduces transitions to other modes. The mode sorter splits the different OAM modes which are then recorded by a CCD camera.}
\label{fig:setup}
\end{figure}

In the last step, after the process, we projectively measure the OAM distribution. To measure the OAM of a light beam, one can use a device called mode sorter, which splits different OAM modes spatially and different regions on a screen can be associated with different values of OAM. Another technique is using a single-mode optical fiber together with a SLM. These optical fibers only couple modes with $\ell = 0$. The SLM is able to change the value of $\ell$ of an incident OAM mode. The value of $\ell$ that the SLM changed in order to couple to the fiber is then equal to the $\ell$ of the OAM mode. A possible way to implement this protocol experimentally as described here is shown in Fig. \ref{fig:setup}.

%%%%%%%%%%%%%%%%%%%%%%%%%%%%%%%%%%%%%%%%%%%%%%%%%%%%%%%%%%%%%%%%%%%%%
%%%%%%%%%%%%%%%%%%%%%%%%%%%%%%%%%%%%%%%%%%%%%%%%%%%%%%%%%%%%%%%%%%%%%%

The information of the input state and the measured output OAM distribution gives us the transition probabilities $p_{\ell' | \ell}$. Together with the thermal probabilities for the corresponding input states we can calculate the joint probabilities $p_{\ell', \ell}$ and, therefore, the work probability distribution $P(W)$ associated with the applied process.
A detailed experimental setup of this protocol as well as an implementation of the Maxwell's Demon using this approach can be found in Ref. \cite{Araujo18}.

%%%%%%%%%%%%%%%%%%%%%%%%%%%%%%%%%%%%%%%%%%%%%%%%%%%%%%%%%%%%%
\vspace*{0.5cm}
\section{Conclusion and Perspectives}
\label{sec:conc}

We have presented an introduction to the new field of quantum termodynamics highlighting the role of experimental investigations based
on all optical setups. The high degree of control of several degrees of freedom of light, like the polarization and the orbital angular momentum, allow testing new theoretical developments like Jarzynki fluctuation relation and realizing proof-of-principle tests of strategies for the interconversion between information and energy, inspired by the Maxwell's demon paradigm. We have also presented novel calculations of work distributions for a quantum harmonic oscillator subjected to squeezing and displacement. These distributions show signatures of their quantum character and we also show that the theoretical results can be experimentally tested using an all-optical scheme. 

We expect that all-optical setups including entangled photons will be used in the near future to investigate quantum effects on fluctuation relations, to implement Maxwell's demon strategies and to investigate the problem of the emergence of the arrow of time.

%%%%%%%%%%%%%%%%%%%%%%%%%%%%%%%%%%%%%%%%%%%%%%%%%%%%%%%%%%%%
\begin{acknowledgments}
The authors would like to thank the Brazilian Agencies CNPq, FAPESC, FAPESP, FAPEG and the Brazilian National Institute of Science and Technology of Quantum Information (INCT/IQ). This study was funded in part by the Coordenação de Aperfeiçoamento de Pessoal de Nível Superior - Brasil (CAPES) - Finance Code 001.
\end{acknowledgments}

%\bibliography{thermo}% Produces the bibliography via BibTeX.

%

\end{document}